# Electrical molecular switch

# addressed by chemical stimuli.

H. Audi,[1] Y. Viero,[2] N. Alwhaibi,[3] Z. Chen,[1] M. Iazykov,[1] A. Heynderickx,[1] F. Xiao,[1]

D. Guérin,[2] C Krzeminski,[2] I.M. Grace,[3] C.J. Lambert,[3,*] O. Siri,[1,*] D.Vuillaume[2,*],

S Lenfant[2,*], H. Klein[1,*]

1) Centre Interdisciplinaire de Nanoscience de Marseille (CINaM), CNRS, Aix Marseille

Université, Marseille, France.

2) Institut d'Electronique, Microélectronique et Nanotechnologie (IEMN), CNRS, Université

de Lille, Avenue Poincaré, Villeneuve d'Ascq, France.

3) Physics Department, Lancaster University, Lancaster, LA1 4YB (UK).

## Abstract

We demonstrate that the conductance switching of benzo-bis(imidazole) molecules upon protonation depends on the lateral functional groups. The protonated H-substituted molecule shows a higher conductance than the neutral one ($G_{pro} > G_{neu}$), while the opposite ($G_{neu} > G_{pro}$) is observed for a molecule laterally functionalized by amino-phenyl groups. These results are demonstrated at various scale lengths : self-assembled monolayer, tiny nanodot-molecule junction and single molecules. From *ab-initio* theoretical calculations, we conclude that for the H-substituted molecule, the result $G_{pro} > G_{neu}$ is correctly explained by a reduction of the LUMO-HOMO gap, while for the amino-phenyl functionnalized molecule, the result $G_{neu} > G_{pro}$ is consistent with a shift of HOMO, which reduces the density of states at the Fermi energy.

# Introduction

Molecular electronic exploits the rich variety of organic molecules to create custom-designed molecular devices for applications in future electronics. The desired function should be encoded in the molecules, which are then connected to electrodes. Among the numerous functional molecules found in the literature, the most striking examples of molecular devices are arguably the molecular switches, i.e. molecules which exist as at least two stable isomers, whose electronic properties can be controllably and reversibly modified by external stimuli.[1] These switches should be distinguished from molecules where a stochastic conductance switching is observed (e.g. for uncontrollable switching driven by the electric field, or electrode/molecule instabilities).[2-6] For conformational switches, the 3D structure of the molecule is modified by the isomerization reaction (stereoisomerization). The cis-trans isomerization induced by light of the azobenzene molecule,[7-12] photoisomerization of diarylethene,[13-16] or hydrogen tautomerization reaction in phtalocyanine,[17] are well-known examples of such switches. In redox switches, the electronic properties depend on the modification of the charge state of the molecules[18-22] through oxidation or reduction depending on the position of the electrochemical potential with respect to the HOMO-LUMO gap of the molecule. It is also possible to control the conductance of molecular junctions by the photo-population of excited states.[23-26] In this case, upon resonant illumination, the electrons photo-injected in the LUMO increase the current though the molecular junctions. In contrast to conformational and redox switches, this effect is not persistent, vanishing rapidly when the excitation is turned off.

Several demonstrations of an alternative approach to electronic conductance modulation by pH control in molecular junctions have been reported.[27-31] For example, the conductance value in the junction can be modified by the protonation or deprotonation of the anchor group of the molecule grafted on a metal surface (electrode). Single molecules of alkanes terminated with diamine or dicarboxylic-acid groups measured by scanning tunneling microscope break-junction (STM-BJ) in a pH-controlled solution show higher conductance at pH = 13 (deprotonated) than at pH = 1 (protonated).[28] For these authors, the basic (pH = 13) or acidic (pH = 1) environment of the molecular junction modifies the chemical specie of the anchoring group: $NH_2$ or $COO-$ in basic environment and $NH_3^+$ or COOH in acidic medium. The $NH_2$ and $COO^-$ species enhance the coupling strength between the molecule and the gold electrode, compared to the $NH_3^+$ and COOH species formed in acidic environment. In the specific case of dithiol terminated alkanes, no significant variation of the conductance with the pH was observed. The same behaviors were also observed for diacid oligo(phenylene ethylene) Langmuir Blodgett films characterized by scanning tunneling microscope (STM)[30] with a higher conductance (ratio $\approx 7$) for the deprotonated form (COO-Au bond at pH = 11.4) than for the protonated form (COOH-Au bond at pH = 5.9). This difference of conductance was interpreted by the chemical modification of the anchoring group with pH as in Ref. 28, but also by the formation of lateral H bonds between neighboring molecules in the films. For these systems, the higher conductance was obtained for the deprotonated form ($G_{depro} > G_{pro}$). More recently, STM-BJ measurements of 4,4'-vinylenedipyridine connected between Ni electrodes showed a conductance switch attributed to change in the electrode/molecule coupling upon protonation, and moreover, the pH required to switch the conductance can also be tuned by the applied electrochemical potential.[32] Protonation and deprotonation of a $\pi$-conjugated molecular system bridged in a molecular junction can also modify the conductance. Oligoaniline derivatives connected into gaps in single walled carbon nanotubes exhibits conductance variations of around one decade between the protonated form (pH = 3 ) and deprotonated form (pH = 11) and this variation was reversible during 5 cycles.[29] The higher conductance was obtained for the protonated form. In another system, a multi-sensitive molecule (pH and light) based on a spyropiran derivative and characterized by STM-BJ shows an increase of the conductance by a factor about 2 after protonation of the spyropiran in the open form.[31] For these π-conjugated molecular systems, the higher conductance was obtained for the protonated form ($G_{pro} > G_{depro}$). We can also mention that an inversion of the rectifying effect in diblock molecular diodes by protonation was observed by STM.[27] This inversion was explained by a modification of the dipolar moment of the molecule with the protonation that induces a reduction of the HOMO and LUMO energies. Consequently, the conduction occurs by a resonant tunneling through the HOMO before the protonation and through the LUMO after. This last result shows that the protonation or deprotonation of the molecule can modify the molecular orbital energy of the molecule addressed in a molecular junction. A similar influence of the protonation on the LUMO position was reported for SAMs, in which the protonation of the terminal carboxylic acid groups converts back and force the molecular junctions between a resistor and a rectifying diode.[33] Protonation has also been used in combination, with light irradiation to block the spiropyran molecule in its merocyanine isomer and avoiding its spontaneous back switching to the spiropyran form.[34] Finally, protonation has been used to induce destructive-quantum-interference in diketopyrrolopyrrole derivative, leading to a reversible decrease of the conductance (1 order of magnitude) upon protonation.[35]

Benzo-imidazole derivatives are molecules known for their sensitivity to pH, exhibiting a different absorption and/or emission signatures depending on their protonation states.[36] Benzo-bis(imidazole) have been recently investigated as precursors of modular fluorescent probes polymers,[37, 38] metal coordination frameworks,[39] or Janus bis(carbene)s and transition metal complexes.[40] They are also very attractive acidochromic systems, because of the presence of two localized π-subunits that can be tuned by reversible protonation to become delocalized.[41] Thus, protonation of benzo-bis(imidazole)s of type **1** ($R^1$ = aniline and $R^2$ = aryl) led to a bathochromic effect ($\Delta\lambda = 13$ nm) that can be explained by the stabilization of the positive charges (color change from yellow to orange).[41] This prompted us to introduce the –SH anchoring group on the benzobis(imidazole) core for grafting such a molecule on gold surfaces in order to evaluate their pH-sensitive switching properties. Here, we demonstrate that the pH effect on the molecule conductance can be controlled by side chain chemistry. A benzo-bis(imidazole) derivative molecule shows a higher conductance in the neutral case ($G_{neu}>G_{pro}$) when laterally functionalized by aniline functions (molecule **A**, scheme 1), while we observe the reverse case ($G_{pro}>G_{neu}$) for the H-substituted molecule (molecule **B**, scheme 1). The molecular conductances were measured at 3 scale lengths : self-assembled monolayers (SAM) on flat Au surface, about hundred molecules grafted on Au nanodots (NMJ : nanodot molecule junction) and single molecule by mechanically controlled break junction (MCBJ). The conductances were measured by C-AFM in the two former cases. The 3 approaches give the same trends. These results are supported by theoretical calculations and we conclude that their opposite behaviors depend on the position of the HOMO resonances relative to the Fermi energy of the electrodes, which modify the density of states at the Fermi energy.

# Experimental section and methods.

## Synthesis of the molecular switch A and B and characterization in solution.

Quinoidal precursors **2** (Bandrowski base)[41] and **3** [42] were reacted with the aldehyde **4** - previously unknown - in EtOH in the presence of piperidine to afford the expected benzobis(imidazoles) **5** and **6**, respectively, in 46 and 49 % yields (Scheme 1). In situ deprotection - using TBAF (tetra-n-butylammonium fluoride) - gave target compounds **A** and **B**, which could be readily reacted in the presence of gold either as molecular junction, or as unit able to form self-assembled monolayers, because of the presence of terminal -SH thiol functions. Details on the synthesis of **4, 5** and **6** and NMR results (Fig. S1) are given in the supporting information. To check the properties of these switches, protonation of **5** (the stable form of **A**, i.e. S-protected) was performed in solution, and clearly showed - as expected - a bathochromic effect (supporting information, figure S1) featuring the formation of the protonated species **7** as model of A•2H$^+$(B•2H$^+$). The protonation of the imidazole site could be clearly identified by color changes[41] and analogy with the related benzimidazole bearing a pyridine unit.[43] Upon protonation, we observed a color change of the solution featuring the protonation of the imidazole and the formation of a delocalized species (as reported in Ref. 41). The lone pair of NH$_2$ units in aniline is indeed less available due to its conjugation with the aromatic ring. As a result, NH$_2$ is then less basic than the benzoimidazole core. This hypothesis is also consistent with Ref. 43, which reports the acid-base properties of 2-pyridylbenzimidazoles. Indeed, the first protonation occurs at the benzimidazole unit and not the pyridine moiety. Since pyridine is more basic than aniline, this observation is in agreement with the protonation of the benzobisimidazole units first in molecule **A**.

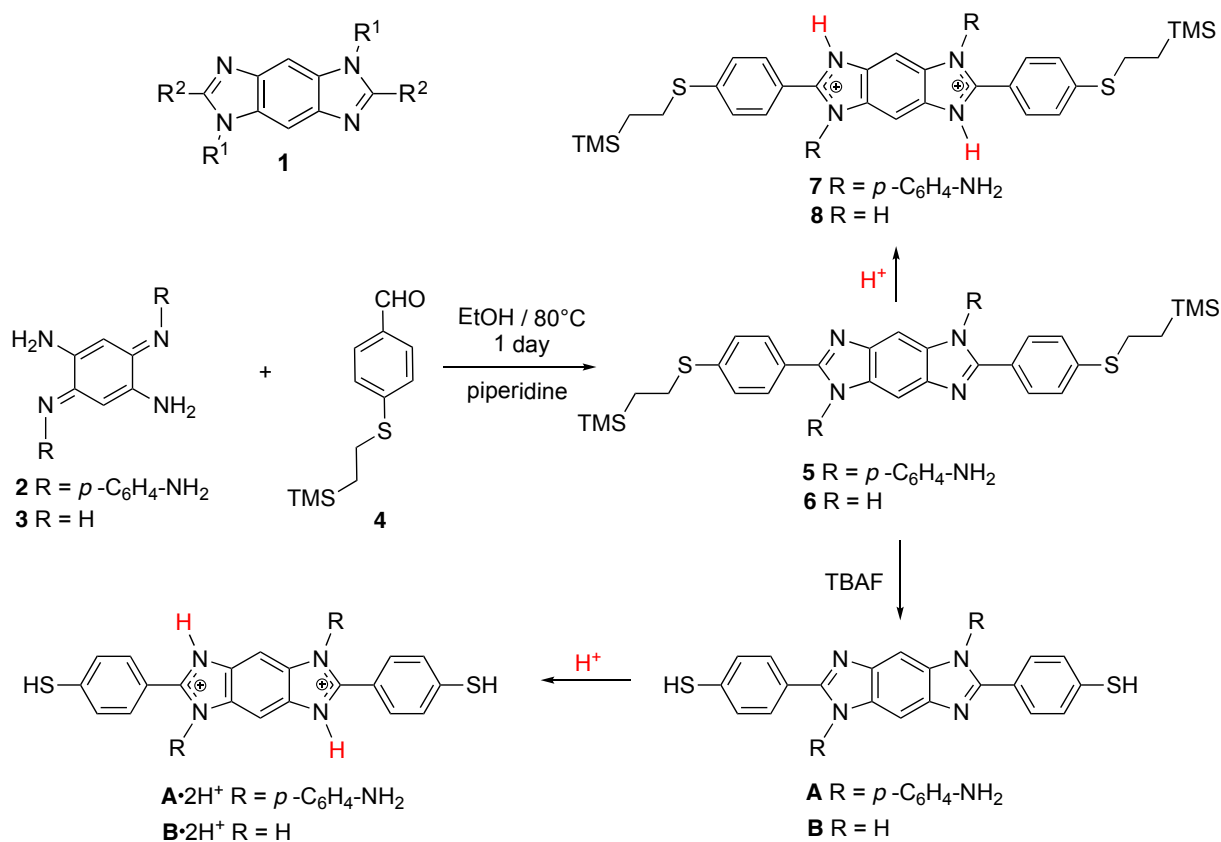

***Scheme 1.*** *Synthesis of benzo-bis(imidazole) devivatives **A** and **B**, and protonation.*

## Preparation of self-assembled monolayers (SAMs) of molecules A and B on gold surfaces.

Glassware was dried in air at 120 °C overnight prior to use. Desilylation and grafting reactions were performed in a nitrogen glovebox. Dimethylsulfoxide (DMSO, anhydrous grade) was stored over freshly activated molecular sieves for 3 days then degassed by nitrogen bubbling for 30 min. Other chemicals were used as received without further purification. Ultra flat gold surfaces were prepared on glass slides following the template stripped gold technique ([TS]Au).[44, 45] 4 eq of tetrabutylammonium fluoride (15 % on alumina, 50 mg for **A**, 60 mg for **B**) were added to 5 mg of protected-thiol **A** or **B** in 20 mL of anhydrous degassed DMSO under stirring. The initial orange solution quickly darkened, then the stirring was continued 1 h under continuous nitrogen bubbling. The medium was quenched by addition of an excess of $CaCl_2.2H_2O$ (20 mg in 1 mL of degassed methanol) to hydrolyze thiolate and precipitate fluorides in excess. After 10 min, the solution was filtered through a 0.45 $\mu$m PTFE filter then immediately used for the preparation of SAMs. The presence of $Bu_4NCl$ adduct and the excess of $CaCl_2$ in the medium were assumed not to interfere in the grafting of the thiol on gold. [TS]Au substrates were immersed for 3 days in the thiol solutions then the functionalized surfaces were cleaned with DMSO, methanol then deionized water. Finally, they were dried under a nitrogen stream.

**Acid/base treatments of the SAMs of A and B.**

Protonation of the SAMs was realized by placing the substrate for 1 min in vapors of diluted hydrochloric acid (i.e. samples held at 2-3 cm above the surface of a 3 M HCl solution). In the same way, conversion back to the neutral form of the molecules in the SAMs was realized by placing the substrate for 1 min in vapors of triethylamine. Each treatment was followed by a short cleaning in ethanol then deionized water.

**Thickness measurements.**

We recorded spectroscopic ellipsometry data (on 1 cm$^2$ samples) in the visible range using a UVISEL (Horiba Jobin Yvon) spectroscopic ellipsometer equipped with DeltaPsi 2 data analysis software. The system acquired a spectrum ranging from 2 to 4.5 eV (corresponding to 300–750 nm) with intervals of 0.1 eV (or 15 nm). Data were taken at an angle of incidence of 70°, and the compensator was set at 45°. We fit the data by a regression analysis to a film-on-substrate model as described by their thickness and their complex refractive indexes. First, a background for the substrate before monolayer deposition was recorded. Secondly, after the monolayer deposition, we used a 2-layer model (substrate/SAM) to fit the measured data and to determine the SAM thickness. We employed the previously measured optical properties of the substrate (background), and we fixed the refractive index of the organic monolayer at 1.50.[46] We note that a change from 1.50 to 1.55 would result in less than a 1 Å error for a thickness less than 30 Å. Overall, we estimated the accuracy of the SAM thickness measurements at ± 2 Å.[47] Albeit other methods (angle resolved XPS, AFM nanoindentation) could be more accurate, ellipsometry has been established as a simple, rapid, and non-destructive tool to estimate the thickness (>1 nm) of SAMs (see a review in Refs. 46, 48).

**XPS measurements.**

XPS was performed with a Physical Electronics 5600 spectrometer fitted in an UHV chamber with a residual pressure of $2 \times 10^{-10}$ Torr. High resolution spectra were recorded with a monochromatic Al Kα X-ray source (hν = 1486.6 eV), a detection angle of 45° as referenced to the sample surface, an analyzer entrance slit width of 400 $\mu$m and with an analyzer pass energy of 12 eV. Semi-quantitative analysis were completed after standard background subtraction according to Shirley's method.[49] Peaks were decomposed by using Voigt functions and a least square minimization procedure with constant Gaussian and Lorentzian broadenings for each component of a given peak.

**Conducting atomic force microscopy (C-AFM).**

Current−voltage characteristics were measured by conducting atomic force microscopy (C-AFM) under a nitrogen flow (Icon, Bruker), using PtIr coated tip (SCM-PIC from Bruker, 0.2 N/m spring constant). To form the molecular junction, the conducting tip was localized at a stationary contact point on the SAM surface at controlled loading force (between 5 to 20 nN).

Current–voltage (I–V) characteristics were acquired with Nanoscope 6 software from Bruker, and treated with Gwyddion v2.44[50] and WsXM v5.0[51] software. Two approaches were used for these C-AFM characterizations: (i) On large flat template-stripped gold surfaces ($^{TS}$Au)[44, 45] modified with SAM, the C-AFM tip is localized at different places on the sample (typically on an array of stationary contact points spaced of 100 nm), at a fixed loading force and the I–V characteristics were acquired directly by varying voltage from 0 V to +200 mV for each contact point. The I-V characteristics were not averaged between successive measurements and typically between 100 and 10,000 I-V measurements were acquired on each sample at different areas of the sample. (ii) On array of ~8 nm (in diameter) single-crystal gold nanodots spaced of 100 nm, fabricated by standard electronic lithography process on (100) Si substrate (resistivity = $10^{-3}$ $\Omega$.cm) and chemically functionalized with molecules, the so-called nanodot molecule junctions (NMJ).[52] The fabrication of the nanodots and the detailed characterization of these nanodot arrays have been reported elsewhere.[53, 54] Images (topography and current) were acquired on the nanodot arrays with a sweep frequency of 0.5 Hz and the voltage applied on the substrate was fixed at 200 mV. From the current images recorded from 1,000 to 4,000 nanodots at a fixed voltage, histograms of the distribution of the current and conductance were obtained for each sample.[52]

**Mechanically controlled break junction (MCBJ).**

Our setup consists of a home-made MCBJ[55, 56] equipped with its dedicated acquisition and control electronics and software. The MCBJ samples (see Fig. S2, supporting information) were made with an Au wire (250 μm diameter, 99.99%, Goodfellow) glued into two quartz capillaries (fused silica, 1mm inner diameter, Vitrocom), which are then glued onto a phosphorous bronze bending beam. An optical glue (NOA61, Nordland) is used for both bondings. The wire is notched in the empty space between the capillaries to initiate the breaking of the junction. The free-standing part of the Au wire is typically 200 µm for our samples, and the unfilled parts of the capillaries act as a reservoir for the organic solutions. To ensure cleanliness of the contacts, samples are broken in pure solvent (dimethyl sulfoxide, DMSO). They are operated for one or two hours at a typical current set-point of 200 pA before measurements begin, to allow mechanical relaxation of the bending beam, and thus stabilization of the electrode distance. We then feed the sample with 10 µL of solution and set a current set-point in the sub nA range. This current imposes the distance between the gold electrodes, and after stabilization, the feedback loop controlling the electrode separation is disabled. We then observe the temporal evolution of the current flowing through the contact. A connection or disconnection of a molecule between the 2 electrodes is expected to induce a sharp change (stepwise) in the current. If we do not observe such events, the feedback is enabled, and the set-point is increased (reducing electrode separation). These operations are repeated until we observe abrupt current jumps (see a typical exemple figure S3, in the supporting information). Current histograms and single molecule conductance are extracted from these measurements as reported in Ref. 56 and detailed in the supporting information.

**Theoretical modeling.**

***Single molecule in gas phase.*** First principles *ab-initio* calculations were performed using the Gaussian software.[57] The geometries were optimized using density functional theory in the framework of the B3LYP functional,[58] 6-311g basis set and the GDIISS optimization algorithm.[59] The influence of pH through proton exchanges at the molecular level is not simple to simulate, due to the interplay of dielectric and solvation effects.[60] Another issue for the modeling is the presence of a counter ion (here Cl-) at a random position, which is needed to equilibrate the total charge of the system.[61] To cope with this issue, we have used a cluster model as defined in Ref. 62 (see the Supporting Information). For each molecule (**A** and **B**) and protonation states, we have calculated the HOMO-LUMO gap, the wave function distributions, the electron affinity (EA) and the ionization potential (IP). The theoretical optical absorption was determined using Time Dependent Functional Density theory calculations (TDDFT).

***Metal/molecule/metal junction.*** To calculate the electron transport through these molecules when a metal/molecule/metal junction is formed, we use a combination of the SIESTA code,[63] which enables the extraction of a Hamiltonian describing the junction and the quantum transport code Gollum.[64] The calculation utilizes a double-$\zeta$ plus polarization orbitals basis set and the GGA[65] exchange correlational functional. To calculate the protonated form, we control the charge on the counter ions (Cl⁻) via the atomic basis set (see Supporting Information). The zero-bias transmission coefficient T(E) is then calculated as a function of the electron energy E. This is then used to evaluate the conductance and electrical current via the Landauer formula.

## Results and discussion.

### Structural characterization of SAMs.

XPS measurements show that the SAMs of molecules **A** and **B** are well formed on the Au surfaces. We observed the peaks of C1s, S2p and N1s with the expected S/C and N/C ratios (see figure S4 and Table S1, Supporting Information). The S2p spectra (for both molecules) showed the contributions of the S-Au bond and the free thiol end-group (Fig. S5, Supporting Information). The N1s spectra showed several peaks (3 for SAM **A**, 2 for SAM **B,** Fig. S6, Supporting Information) attributed to the different chemical environments of the N atoms (see discussion, Supporting Information). Table 1 presents the thickness of SAMs **A** and **B** measured by ellipsometry before and after the acid/base treatments. The theoretical length of molecules **A** and **B** are estimated to 1.9 nm (see geometry optimization, Supporting Information). It appears that SAM **B** (thickness ~ 1.9 nm) is denser and better organized than SAM **A** (thickness ~ 1.3 nm, thus an average tilt angle of the molecule of ~ 50° with respect to the surface normal). This difference can be explained by the more planar shape of molecule **B**, which is more favorable to a compact stacking. The treatments with vapors of hydrochloric acid and triethylamine do not induce a significant variation of the thickness of SAMs **A** and B, indicating a good stability.

| | pristine | HCl exposure | NEt$_3$ exposure |
|---|---|---|---|
| **SAM A** | 1.3 ± 0.2 | 1.3 ± 0.2 | 1.5 ± 0.2 |
| **SAM B** | 1.9 ± 0.2 | 1.9 ± 0.2 | 1.8 ± 0.2 |

***Table 1.*** *Thickness (in nm) measured by ellipsometry for SAMs **A** and **B** before (pristine) and after the acid and base treatments. Average values from 2-3 samples and 3 measurements at different places on each samples.*

**Electrical characterization.**

***SAM on flat $^{TS}$Au surfaces characterized by C-AFM.***

We measured the electron transport properties in molecular junctions, which consist of SAMs on flat $^{TS}$Au electrodes, and contacted by C-AFM. For this purpose, the loading force of the C-AFM tip is kept very low (around 5 nN) in order to avoid any mechanical deformation of the SAM (< 0.1 nm) and consequently any modification of its electronic properties.[66-68] The I-V curves were accumulated over different zones (typically 4) of the sample, and after removing the short-circuited curves (corresponding to curves with current reaching compliance saturation of the C-AFM preamplifier at very low voltage around few mV) we end-up with $10^3$ to $10^4$ I-V curves (see I-V curves in Fig. S7, Supporting Information). The current histograms were constructed from these curves at a fixed voltage (35, 70, 140 and 200 mV) for SAMs after fabrication (pristine), after protonation and after conversion back to the neutral state (Figure 1 at 200 mV, other data at lower voltages in the Supporting Information, Figs. S8 and S9) according to the procedure described in the experimental section. These current histograms were well fitted by one log-normal distribution (parameters given in figure caption and table S2, Supporting Information).

For both SAMs of molecules **A** and **B**, the current distributions are shifted by the protonation and the back conversion to the neutral state of the molecules, and we observe that the mean current for the pristine SAM has an intermediate value between the mean currents measured on the protonated and neutral SAMs. Specifically for the molecule **A** (Figure 1-a), the mean current initially measured at $4.8 \times 10^{-9}$A for the pristine SAM is between the mean current of the protonated SAM of $1.5 \times 10^{-9}$A and the neutral SAM at $1.4 \times 10^{-8}$A. For molecule **B**, the mean currents for the pristine, protonated and neutral SAMs are measured at $8.3 \times 10^{-10}$ A, $2.0 \times 10^{-9}$A and $3.2 \times 10^{-10}$A respectively (Figure 1-b). It is likely that in the "pristine" case, just after the SAM grafting, there is a mix of neutral and protonated molecules. Thus, we observed that the molecule **A** is more conducting in the neutral state, while the molecule **B** exhibits a higher conductance in the protonated state. The ratios of the mean currents are $I_{neu}(A)/I_{pro}(A)=9.3$ and $I_{pro}(B)/I_{neu}(B)=6.3$ for molecule **A** and **B**, respectively (table 2).

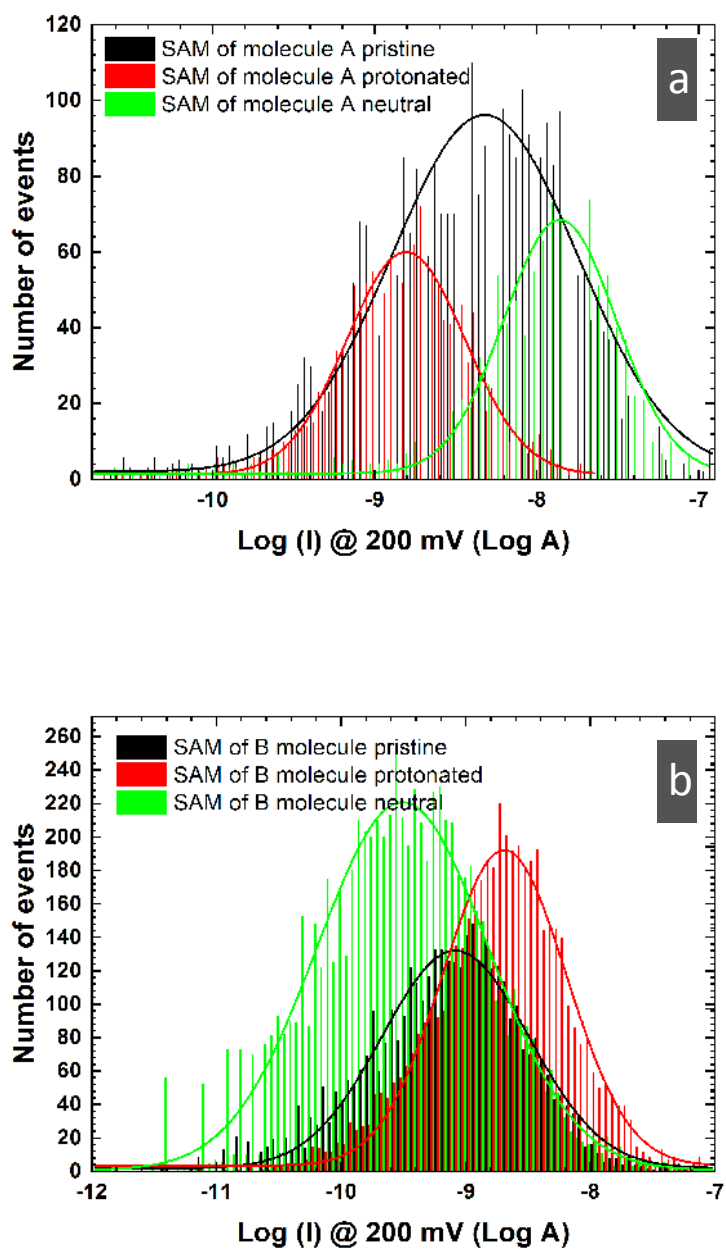

**Figure 1.** *Histograms of the current at a fixed bias of 200 mV for (a) SAM of molecule **A** after fabrication (3698 junctions) in black (log μ = -8.32 (4,8x10⁻⁹ A), log σ = 0.57), after protonation in red (1588 junctions, log μ = -8.81 (1.5x10⁻⁹ A), log σ = 0.36) and after converted back to the neutral state in green (1186 junctions, log μ = -7.86 (1.4x10⁻⁸ A), log σ = 0.69); (b) SAM of molecule **B** after fabrication (6096 junctions, log μ = -9.08 (8.3x10⁻¹⁰ A), log σ = 0.56) in black, after protonation in red (9338 junctions, log μ = -8.69 (2x10⁻⁹ A), log σ*

= 0.50) and after converted back to the neutral state in green (5034 junctions, log μ = -9.50 (3.2x10⁻¹⁰ A), log σ = 0.68). Solid lines are the log-normal distributions fitted on the data.

These current histograms were measured for successive cycles of protonation and back conversion to the neutral state for the two molecular junctions. The parameters (mean current and standard deviation) of the fitted log-normal distributions for these histograms are plotted in figure 2. The opposite behaviors of the conductance variations upon protonated/ neutral is reproduced for 3 cycles albeit we observe a progressive decrease of the current ratios.

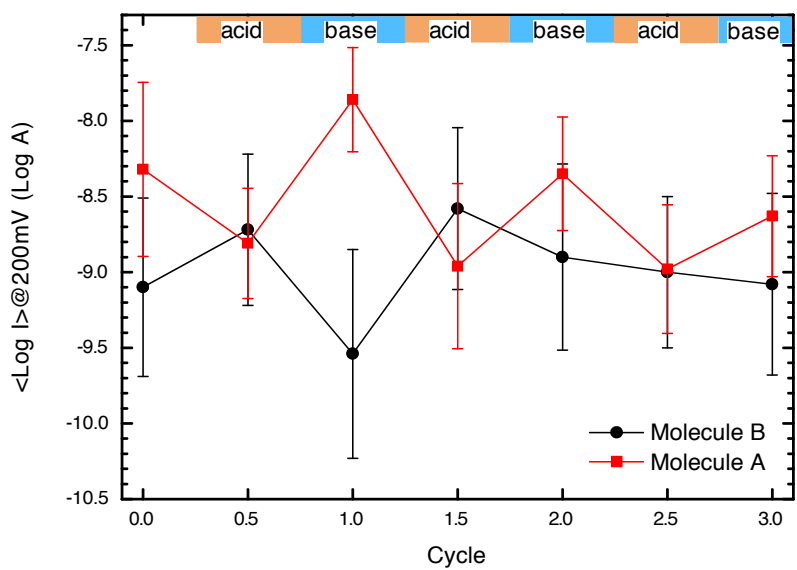

***Figure 2.*** *Plot of the log mean current and standard deviation (plotted as error bar) of the fitted log-normal distributions for the current histograms measured by C-AFM at 200 mV for the two molecules and for 3 successive cycles of protonation (acid) and back conversion to the neutral form (basic).*

For molecules **A**, the ratios $I_{neu}(A)/I_{pro}(A)$ are 9.3, 4.1 and 2.2. For molecule **B**, we get $I_{pro}(B)/I_{neu}(B)$ ratios of 6.3, 2.1 and 1.2 for the three cycles. This "fatigue" effect, usually observed in molecular switches (optical,[69] redox[22] or pH[29]), may have several causes (molecule desorption or degradation, uncontrolled chemical reactions,…) and strategies have been recently demonstrated (e.g. mixed SAMs with alkylthiols) to maximize the conductance switching and reduce this fatigue effect.[70, 71]

| | MCBJ | NMJ | SAM/Au$^{TS}$ | theory |
|---|---|---|---|---|
| A ($G_{neu}/G_{pro}$) | n.m. | 3.4 | 9.3 | 6 |
| B ($G_{pro}/G_{neu}$) | 4.5-6.5 | 2.5 | 6.3 | 15 |

**Table 2**. *Switching ratios for molecules **A** and **B** measured at 3 scale lengths by 3 approaches, and theoretical calculations (n.m. :not measured)*

**Molecules grafted on sub-10 nm nanodots (NMJ) and characterized by C-AFM.**

The changes in the molecular junction conductance for the protonated and neutral states of the molecules were also assessed with another platform of smaller size, made from hundreds of molecules grafted onto tiny crystalline Au nanodots and contacted by C-AFM (NMJ).[52, 53] The current is measured by scanning the array of NMJs, with a C-AFM tip at a given bias applied on the substrate. A current is measured only when the tip is localized on the molecular junction (Fig. 3).[52] The current through the native $SiO_2$ covering the space between the nanodots is below the detection limit of our apparatus. C-AFM images on this array of nanodot/molecules junctions taken at + 200 mV for both the molecules and for the two states (neutral and protonated) are shown in the supporting information (Fig. S10). From these C-AFM images and with help of the Gwyddion software,[50] we extract a value of the current for each electrically active nanodot/molecules junctions in each C-AFM image, and we construct current histograms.[52]

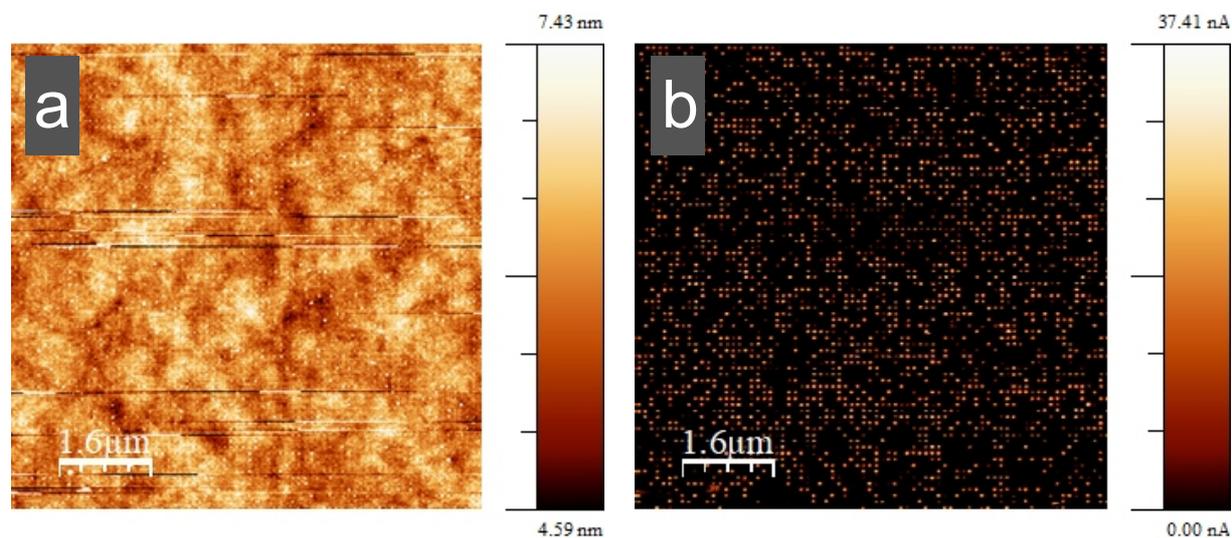

**Figure 3.** *Example of* **(a)** *topography image (nanodots are bright spots) and* **(b)** *current image (bright spots are the current value of each NMJ) obtained simultaneously on NMJs of neutral molecules* **A** *and measured by C-AFM (loading force10 nN and $V_{bias}$ = 200 mV). 2456 NMJs are measured.*

The current histograms at 200 mV for both the molecules and in both states (protonated and neutral) are presented in figure 4. These histograms are well fitted by log-normal distributions (solid lines) and the fitted parameters (log mean current, log $\mu$, and log standard deviation, log $\sigma$) are given in table S2 (Supporting Information) and in the figure captions. We clearly observe a shift of the distribution between the protonated and neutral NMJs, with higher currents for the neutral molecule **A** ($G_{neu}(A)>G_{pro}(A)$) and a decrease of the current for the neutral molecule **B** ($G_{pro}(B)>G_{neu}(B)$). We define $I_{pro}$ and $I_{neu}$ as the mean current value after protonation and after converted back to the neutral state, respectively, we obtain $I_{pro}(A)=8.5x10^{-9}A$ and $I_{pro}(B)=4.2x10^{-8}A$ for molecule **A** and **B**, respectively, and $I_{neu}(A)=2.9\text{x}10^{-8}A$ and $I_{neu}(B)=1.7\text{x}10^{-8}A$ for molecule **A** and **B**, respectively. These current values lead to ratios of conductances $I_{neu}(A)/I_{pro}(A)=3.4$ and $I_{pro}(B)/I_{neu}(B)=2.5$ for molecule **A** and **B**, respectively, a little bit smaller then on $^{TS}Au$ surfaces (table 2). This inversion of the ratios with the nature of the side chains of the molecule will be further discussed in the theoretical section below.

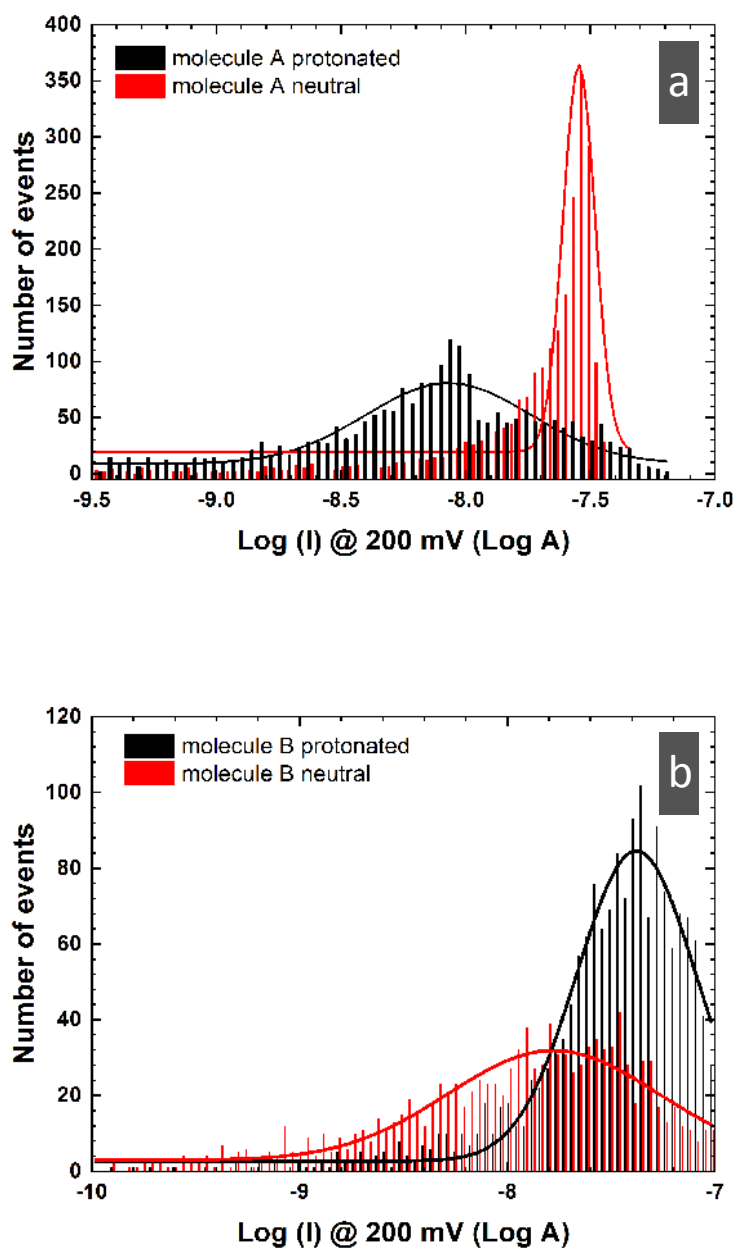

**Figure 4.** *Histograms of the current at a fixed bias (200 mV) measured by C-AFM on **(a)** array of NMJs of molecule **A** after protonation in black (2445 junctions, log μ = -8.07 (8.5x10⁻⁹ A), log σ = 0.33) and for neutral molecules in red (2456 junctions, log μ = -7.54 (2.9x10⁻⁸ A), log σ = 0.05); **(b)** NMJs of molecule **B** after protonation in black (1634 junctions, log μ = -7.38 (4.2x10⁻⁸ A), log σ = 0.29) and for neutral molecules in red (1154*

*junctions, log μ = -7.78 (1.7x10$^{-8}$ A), log σ = 0.51). Solid lines are the log-normal distributions fitted on the data.*

### Single molecule conductance measurements by MCBJ.

Finally, we compare these results with single molecule experiments using the MCBJ technique. Following the protocol described in the experimental section, we performed single molecule conductance measurements on molecules **A** and **B**, starting with a millimolar solution of the neutral molecules in DMSO. However, we never obtained stable molecular junctions for the molecule **A**, which we attribute to steric hindrance caused by the bulkier side groups. Therefore, only results for molecule **B** are reported here. Figure 5 shows the histograms constructed from current recordings at $V_{bias}$ = 35 mV (a typical current vs. time trace is given figure S11 in the supporting information) for a MCBJ operated in pure DMSO, in a millimolar solution of neutral and protonated molecule **B**. While the histogram of currents for the molecule **B** exhibits a series of evenly spaced current peaks, the measurement performed in pure DMSO does not exhibit such features. Moreover, the current histogram after protonation is shifted to higher currents. Measurements are realized for bias voltage of 35, 70 and 140 mV following the same protocol (for comparison with the C-AFM results). Similar current histograms are obtained (see figures S11-S13 for $V_{bias}$ = 35, 70 and 140 mV, Supporting Information). Measurements at higher bias voltages (i.e. 200 mV) are not reported here, because molecular junctions become unstable above 140 mV. These instabilities are attributed to electromigration processes in the metallic electrodes, and current-induced heating of the molecular junction.[72]

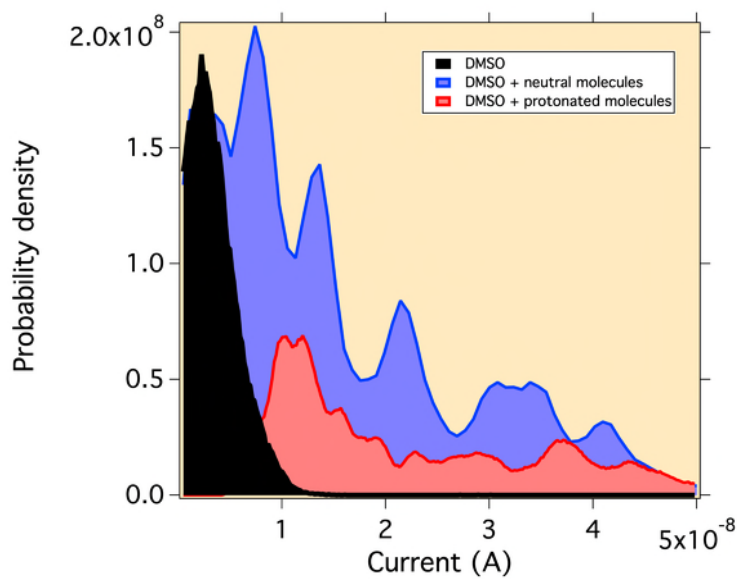

***Figure 5****. Histograms of currents (at V_{bias} = 35 mV) measured with the MCBJ in pure DMSO (dark) and in a 1 mM solution of molecules **B** in the neutral state (blue) and protonated state (red). The current histogram after protonation is shifted to higher currents, see text and Figs. 6 and S17 (in the supporting information) for the data analysis (multi-peak) and the single molecule conductance determination.*

We obtained complex conductance histograms, which prevent us from a straightforward analysis. We thus developed an automated analysis protocol (see more details in the supporting information). In brief, from the current histograms converted to conductance, we calculate the differences between the most probable current values (peaks in Fig. 5). If these differences correspond to connection/disconnection of molecules in the contact, they should share a common multiple, the conductance of a single molecule. We plot in figure 6 the peaks of conductance versus a discrete number of molecules $N_{MOL}$. Arbitrarily, the lowest conductance value (lowest current peak) is attributed to $N_{MOL} = 1$, since we do not observe current jumps of lower amplitude. These plots clearly show that for all the measurements, the conductance values are fitted by a linear function $f(N_{MOL}) = G_{MOL} N_{MOL}$, $G_{MOL}$ being the conductance of a single molecule. We note that there is some missing points (i.e. peaks), which reflects the stochastic nature of these molecular junctions. The conductance values are expressed in the conductance quantum units, $G_0 = 2e^2/h$ (7.75x10$^{-5}$ S), with e the electron charge and h the Planck constant. The extracted $G_{MOL}$ values are reported in table 3. We note that the conductances vary with the applied voltage, while the conductances of the SAMs (see Table S3, Supporting Information) are not dependent on the voltage (linear I-V regime < 200 mV). However, conductances of single molecules is very sensitive to the atomic details of the molecule/electrode contact. For example, it has been reported[73, 74] that the conductance depends whether the molecule is connected on a hollow site of the gold surface or to a gold ad-atom (this variability being averaged in a larger area junction).[75]

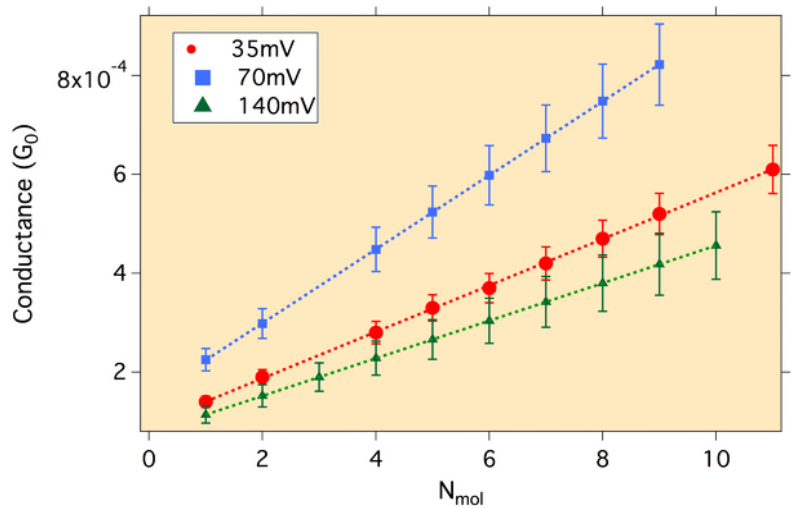

*Figure 6*. *Conductance values of the peaks observed in the histograms of Fig. 5 (and figures S9-S11) versus a number of molecules (neutral molecule **B**). A linear fit gives the $G_{MOL}$ values reported in table 2 for the 3 applied voltages. Conductance values are given in conductance quantum units, $G_0$.*

| Bias (mV) | $G_{MOL}$ ($G_0$) neutral | $G_{MOL}$ ($G_0$) protonated |
|---|---|---|
| 35 | $4.7 \pm 0.2 \times 10^{-5}$ | $3.40 \pm 0.02 \times 10^{-4}$ |
| 70 | $7.50 \pm 0.03 \times 10^{-5}$ | $4.80 \pm 0.46 \times 10^{-4}$ |
| 140 | $3.80 \pm 0.01 \times 10^{-5}$ | $1.72 \pm 0.33 \times 10^{-4}$ |

***Table 3 :*** *Values of the molecule conductance, Gmol, at several voltages and in the neutral (from data Fig. 6) and protonated forms (data Fig S17 in the supporting information).*

We have then applied the same measurement and analysis protocol to the protonated molecule **B** (Figs. S14-S16 in the supporting information). The protonation is operated in-situ by exposing the junction with molecules to HCl vapors for one minute. After the exposition, the base tunneling current in the junction, for given electrode separation, increases by approximately one order of magnitude. This current increase is attributed to the dissolution of acid in the solvent, leading to the opening of an ionic current channel in the junction. One of the main benefits of the measurement and analysis method used here, lies in eliminating the constant (or slowly varying) tunneling or ionic contribution to the net conductance.[56] In these conditions, it makes sense to compare conductance values of protonated and neutral molecules. Table 3 gathers the conductance values obtained for neutral and protonated molecules **B** at $V_{bias} = 35, 70$ and 140 mV (see supporting information, figures S14-S16 given the corresponding current traces and histograms of currents and figure S17 for the $G_{MOL}$ vs. $N_{MOL}$ plots). In all cases, the conductance of protonated molecules is higher than the conductance of the neutral ones. From these values, we can calculate a conductance ratio between protonated and neutral molecule ranging from 4.5 to 6.5 (table 2). These results show that we have operated a molecular switch, using a chemical stimulus, at the scale of a single molecule.

*Modeling molecules in gas phase*

We first calculated the optimized geometries, the electron affinity (EA), the ionization potential (IP) and the HOMO-LUMO gap for molecules **A** and **B** in the neutral and protonated forms (Figs. S18-S20, Supporting Information). For molecule **A**, we considered the $4H^+$ and $2H^+$ cases, and we find that the HOMO-LUMO gap is the same in the two cases (Fig. S19, Supporting Information). Basically, the protonation tends to reduce the HOMO-LUMO gap of the molecules and increases the IP and EA. Both the HOMO and the LUMO of molecules **A** and **B** are characterized by highly delocalized wave functions. The protonation of the nitrogen atoms strongly modify the nature of the HOMO as it becomes strongly localized on the protonation sites i.e. on the chloride counter-ion. On the contrary, the LUMO wave functions remain strongly delocalized on the aromatic cycles.

*Modeling electron transport in metal/molecule/metal junctions*

To explain this switching behavior, we calculated electron transport through the two molecules when attached to gold electrodes. The transmission curves T(E) in Figure 7 show the gap between the HOMO and LUMO resonances is approximately 2eV. To perform a comparison between molecule **A** and **B**, we first calculate their transport properties for identical contacting geometries as shown in Figure 7. Molecule **A** in the neutral form shows that the DFT predicted Fermi energy ($E_F^0$) lies close to the HOMO resonance and gives a conductance value of $0.06G_0$. Upon protonation, the HOMO-LUMO gap is decreased and the position of the Fermi energy lies in the middle of the gap. The value of conductance decreases to $0.01G_0$ in agreement with trend shown in the SAM measurement of figures 1 and 4. The thickness of the SAM ~ 1.3nm compared to the molecule length (1.8nm) means the molecule forms a more tilted geometry in the SAM. However, we find that the general trend upon protonation is not dependent on the contact angle geometry (figure S24, Supporting Information). Also, electron transport though tilted molecule A contacting the gold electrode though the $NH_2$ anchor group can be discarded, since the conductance is 3 orders of magnitude lower (Fig. S25, Supporting Information).

For molecule **B**, we find that the DFT calculation positions the Fermi energy close to the HOMO, leading to a similar transmission for Molecule **A** (Figure S22, Supporting Information). However, the calculated IP of molecule **B** is larger than **A** (meaning the HOMO is at a lower energy). Figure S19 (Supporting Information) shows this difference to be ~ 0.5eV, and therefore we shift the Fermi energy by this amount and define $E_F^1 = E_F^0 + 0.5eV$ (Fig. 7b). The value of the conductance for the neutral case at $E_F^1$ is $0.002G_0$. Protonation again decreases the HOMO-LUMO gap, but now the transmission is higher at $E - E_F^1 = 0eV$ and the conductance now increases to a value of $0.03G_0$, again in agreement with the measured trend (Figs. 1, 4 and 6). Also, the predicted conductance is much larger than the measured value of G (Table 3), which can be attributed to the underestimation of the HOMO-LUMO gap. However the conductance ratios for molecular **A** $G_{neu}/G_{pro} = 6$ and $G_{pro}/G_{neu} = 15$ for molecule **B** are in agreement with our measured ratios (Table 2). Thus, the changing behavior on protonation between molecule **A** and **B** can be attributed to the difference in the relative position of the HOMO resonance with respect to the Fermi energy.

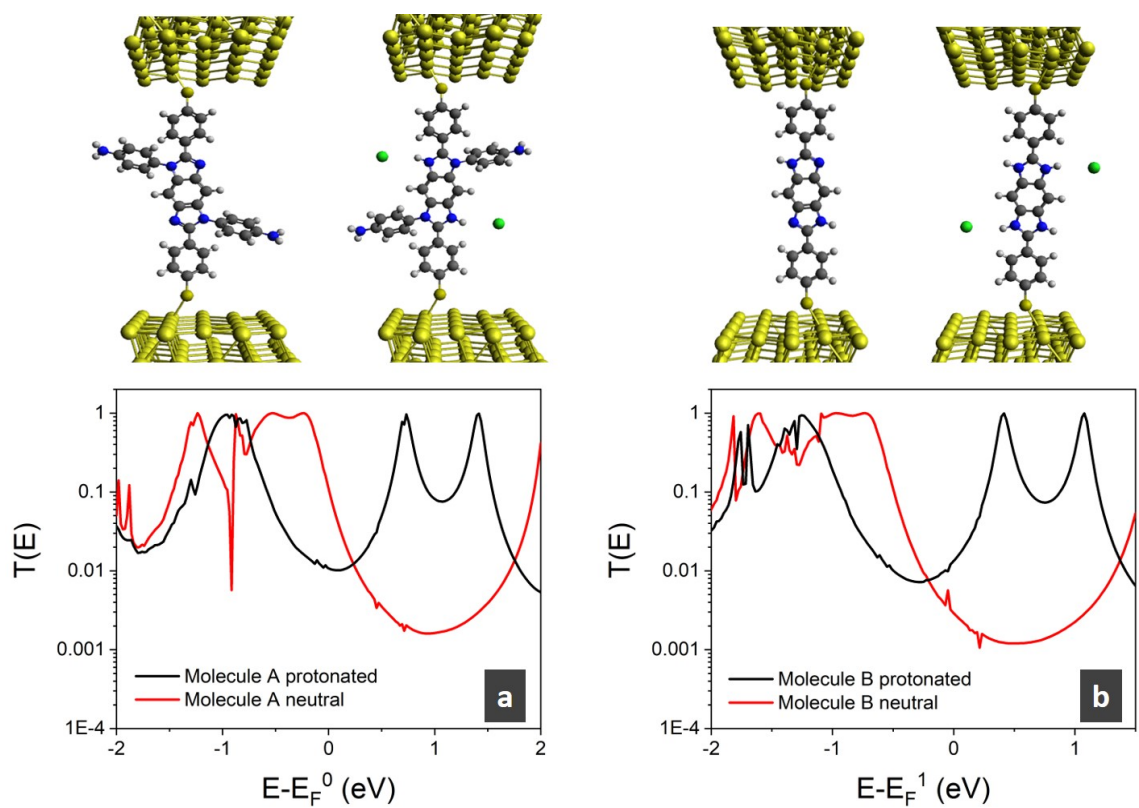

***Figure 7***. *Zero bias transmission coefficient T(E) for molecule **A (a)** and molecule **B (b)** in a linear geometry for the protonated (black lines) and neutral (red lines) forms.*

## Conclusion.

A multi-scale characterization of the electron transport though molecular junction upon protanation demonstrates that their conductance switching depends on the lateral functional groups. The molecular conductances were measured at 3 scale lengths : self-assembled monolayers on flat Au surfaces, about hundred molecules grafted on Au nanodots (measured by C-AFM in these two cases) and single molecules by mechanically controlled break junction, and. These 3 approaches demonstrate that the effect of pH modulation on the molecule conductance can be controlled by side chain chemistry. A benzo-bis(imidazole) molecule shows a higher conductance in the neutral state ($G_{neu} > G_{pro}$) when laterally functionalized by amino-phenyl groups, while we observe the reverse case ($G_{pro} > G_{neu}$) for the H-substituted molecule. These results are understood with the help of theoretical calculations which attribute the different behaviors to the position of the HOMO resonances relative to the Fermi energy of the electrodes.

## Associated content

### *Supporting Information*

The Supporting Information is available free of charge on website at DOI: xxxxx

Synthesis, NMR, UV-vis spectra, detailed protocols MCBJ measurements, XPS spectra, detailed current vs. time traces during MCBJ measurements, C-AFM images on NMJs, C-AFM current-voltage curves on [TS]AU/SAM, calculated optimized geometries, calculated electronic structures, calculated molecular orbitals, calculated optical absorption curves and calculated electron transport in metal/molecule/metal junctions.

## Author information

### *Author contributions*

H.A., Z.C., A.H., F.X. and O.S. synthesized the molecules, optimized the procedures, and characterized the compounds by spectroscopy (NMR and UV-vis measurements). Y.V., D.G. and S.L. fabricated the devices on gold nano dots and flat gold surfaces and carried out the C-AFM measurements. D.G. performed the XPS measurements. M.I. and H.K. did the single molecule conductance measurements. C.K. performed the theoretical calculations for molecules in gas phase. N.A., I.M.G and C.J.L carried out the electron transport calculations of the molecular junctions. O.S., D.V. and H.K. conceived the experiments and supervised the project. D.V., O.S., D.G., C.K., S.L., I.M.G, C.J.L. and H.K. wrote the paper with the help and contributions of all the authors.

*Corresponding authors*

\* olivier.siri@univ-amu.fr

\* klein@cinam.univ-mrs.fr

\* c.lambert@lancaster.ac.uk

\* dominique.vuillaume@iemn.fr

\* stephane.lenfant@iemn.fr

*ORCID*

H. Audi: 0000-0001-9291-9409

A. Heynderickx: 0000-0002-4991-6592

I.M. Grace: 0000-0001-8686-2290

C.J. Lambert: 0000-0003-2332-9610

O. Siri: 0000-0001-9747-3813

D.Vuillaume: 0000-0002-3362-1669

S Lenfant: 0000-0002-6857-8752

H. Klein: 0000-0002-8625-4778

*Notes*

The authors declare no competing financial interest.

## Acknowledgments

This work was funded by the ANR grant FOST (ANR-12-BS10-01801). We thank Dominique Deresmes (IEMN) for help and advices (C-AFM measurements). Support from the UK EPSRC is acknowledged, through grant nos. EP/N017188/1, EP/M014452/1, EP/P027156/1 and EP/N03337X/1. Support from the European Commission is provided by the FET Open project 767187 – QuIET.

**Supporting information**

# Electrical molecular switch addressed by chemical stimuli

*H. Audi,[1] Y. Viero,[2] N. Alwhaibi,[3] Z. Chen,[1] M. Iazykov,[1] A. Heynderickx,[1] F. Xiao,[1]*

*D. Guérin,[2] C Krzeminski,[2] I.M. Grace,[3] C.J. Lambert,[3,*] O. Siri,[1,*] D.Vuillaume[2,*],*

*S Lenfant[2,*], H. Klein[1,*]*

1)  Centre Interdisciplinaire de Nanoscience de Marseille (CINaM), CNRS, Aix Marseille Université, Marseille, France.

2) Institut d'Electronique, Microélectronique et Nanotechnologie (IEMN), CNRS, Université de Lille, Avenue Poincaré, Villeneuve d'Ascq, France.

3) Physics Department, Lancaster University, Lancaster, LA1 4YB (UK).

# I. Synthesis.

- *Synthesis of **4**.*

Step 1:

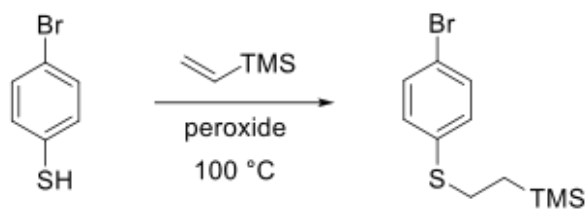

The target compound was prepared as reported[1-2] using a modified procedure. To a Schlenk tube was added 6.9 g (36.49 mmol) of 4-bromothiophenol, 6.6 ml (42.82 mmol) of vinyltrimethylsilane, and 0.95 ml (5.05 mmol) tert-butyl peroxide under argon. The reaction mixture was stirred at 100°C for 10 h. The reaction mixture was diluted by adding 200 ml of hexane, and the obtained solution was washed once with a 10% sodium hydroxide aqueous solution. The organic layer was separated, dried (MgSO$_4$), and then concentrated under reduced pressure. The crude product was purified by column chromatography over silica gel (eluent: cyclohexane) to afford the target protected thiol as a yellow oil (m = 8.6 g, 83% yield).

[1]H NMR (400 MHz, CDCl$_3$): δ(ppm) 7.38 (d, J = 8.6 Hz, 2H), 7.15 (d, J = 8.5 Hz, 2H), 2.96 – 2.88 (m, 2H), 0.93 – 0.87 (m, 2H), 0.03 (s, 9H).

Step 2 :

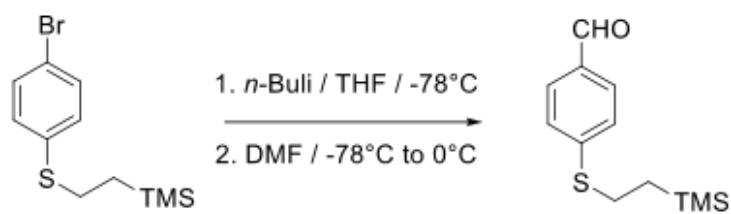

A solution of 2-(trimethylsilyl)ethyl-4-bromophenyl sulfide (m = 1,0 g, 3.46 mmol) in 30 ml of dry THF was treated with n-BuLi (v = 1.7 ml of a 2.5M solution in hexane, 4.25 mmol) dropwise under argon at -78 °C. After stirring for 1h, the reaction mixture was quenched with anhydrous dimethylformamide (v = 0.33 ml, 4.15 mmol) at -78 °C. The mixture was then further stirred for an hour at low temperature, an additional hour at room temperature, and poured into aqueous $NH_4Cl$. The mixture was extracted with $CH_2Cl_2$ (2 × 30 ml). The combined organic layers were washed with water, dried ($MgSO_4$) and filtered. The solvent was evaporated under reduced pressure, and the crude product was purified by column chromatography over silica gel (eluent: $CH_2Cl_2$: PE=3:1) to afford **4** (m = 0.60 g, 73% yield) as a colorless oil.

[1]H NMR (400 MHz, $CDCl_3$): δ(ppm) 9.90 (s, 1H), 7.74 (d, J = 8.3 Hz, 2H), 7.31 (d, J = 8.3 Hz, 2H), 3.12 – 2.93 (m, 2H), 1.04 – 0.87 (m, 2H), 0.06 (s, 9H). [13]C NMR ($CDCl_3$): δ(ppm) 191.46, 147.62, 133.22, 130.20, 126.45, 77.55, 77.23, 76.91, 28.00, 16.43, -1.56.

*- Synthesis of **5**.*

To a solution of **4** (m = 113 mg, 0.47 mmol) in 3ml of ethanol was added the Bandrowski base **2** (m = 70 mg, 0.22 mmol) and two drops of piperidine. The mixture was refluxed for 24h. The resulting precipitate was filtered, washed with ethanol and dried to give the desired product **5** as beige solid ( m = 65 mg, 43% yield).

[1]H NMR (250 MHz, DMSO-$d_6$): δ(ppm) 0.04 (s, 18H); 0.83-0.90 (m, 4H); 2.98-3.05 (m, 4H); 5.50 (s, 4H); 6.71-6.75 (4H, d, J = 8.5 Hz); 7.05-7.09 (2H, d, J = 8.5 Hz); 7.23-7.26 (m, 6H); 7.52-7.56 (4H, d, J = 8.25 Hz). ESI-HRMS: Calculated for $C_{42}H_{48}N_6S_2Si_2$, 756.2920 ; Found 757.2989 [M+H]$^+$.

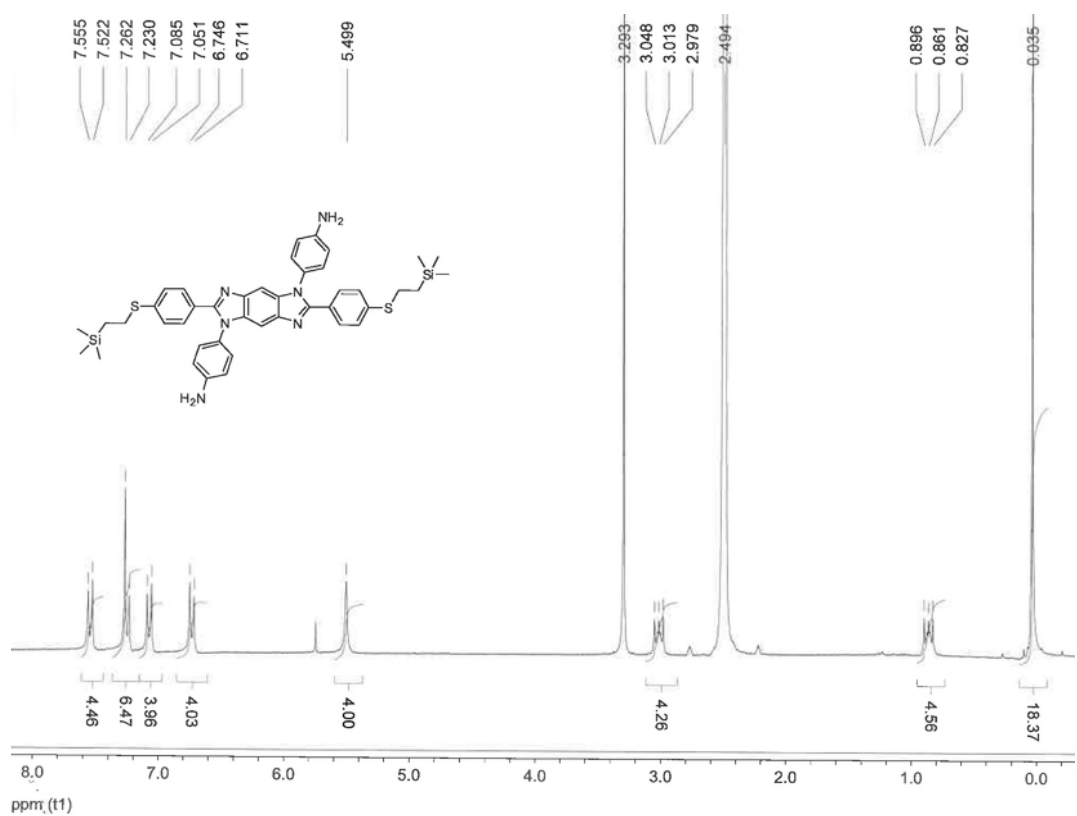

***Figure S1-a***. *$^1H$ NMR spectrum of **5** in DMSO.*

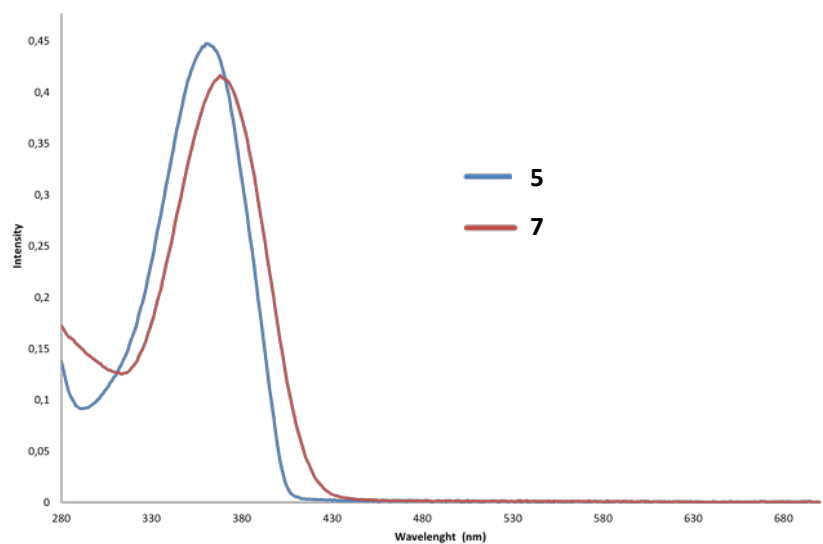

***Figure S1-b.*** *UV-vis absorption of **5** (C = (1.88x10⁻⁵M), λ_max = 360 nm) and its protonated form **7** (λ_max = 368 nm) in THF.*

*- Synthesis of **6**.*

To a solution of **4** (m = 100 mg, 0.73 mmol) in 3ml of ethanol was added reagent **3** (m = 87.6 mg, 0.37 mmol) and three drops of piperidine. The mixture was refluxed for 24h. The resulting precipitate was filtered, washed with ethanol and dried to afford **6** as beige solid (m = 105 mg; 49%).

[1]H NMR (400 MHz, DMSO-d$_6$): δ(ppm) 12.59 (d, J = 17.7 Hz, 2H), 8.04 (d, J = 12.9 Hz, 4H), 7.76 (s, 1H), 7.59 (s, 1H), 7.37 (d, J =13.6 Hz, 4H), 3.08 – 2.97 (m, 4H), 0.92 – 0.78 (m, 4H), 0.00 (s, 18H). ESI-HRMS: Calculated for C$_{30}$H$_{38}$N$_4$S$_2$Si$_2$, 574.2076 ; Found 575.2147 [M+H]$^+$.

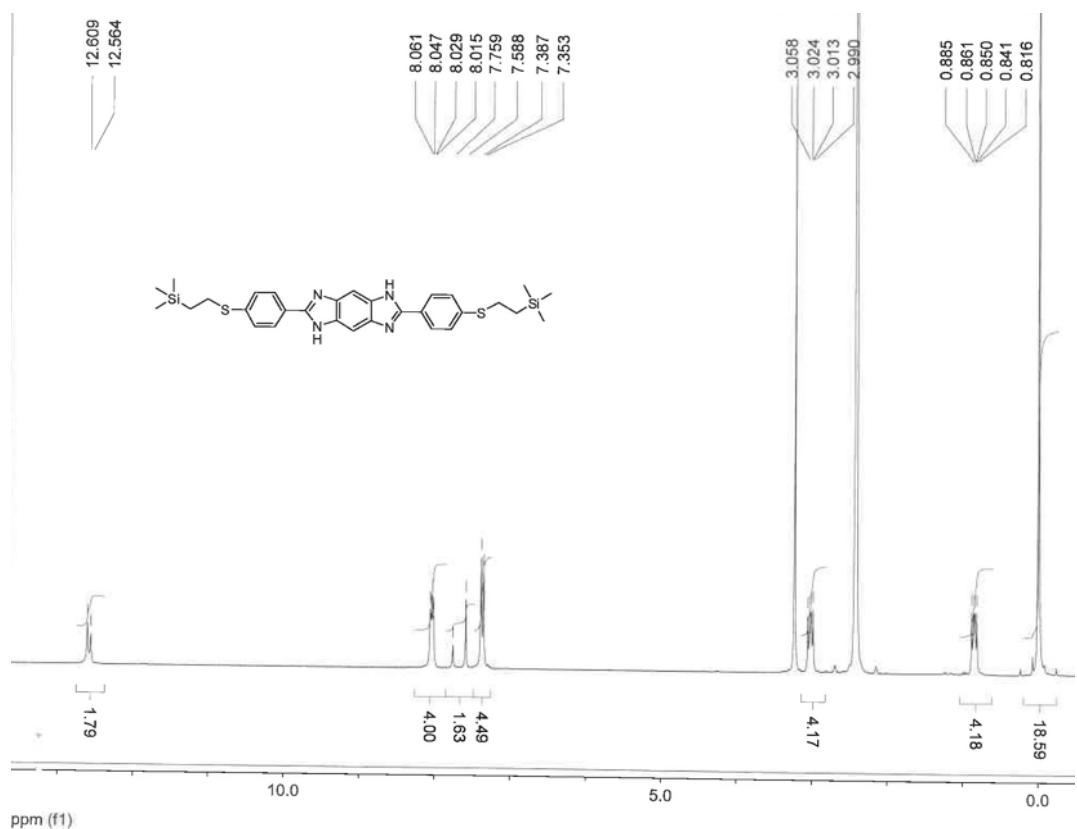

***Figure S1-c****. [1]H-NMR spectrum of **6** in DMSO.*

## II. Mechanically Controlled Break Junction (MCBJ).

***MCBJ set-up.***

The MCBJ, as introduced by Muller et al.[3] consists in stretching a metallic wire by bending an elastic substrate supporting the wire with a mechanical actuator until the junction breaks, giving two separate electrodes. The electrode separation can then be adjusted by a feedback loop using the current flowing through the junction. MCBJ exhibits excellent mechanical stability, which results from the reduction of the mechanical loop connecting one electrode to the other. For this study, our MCBJ is operated at room temperature in organic solutions.

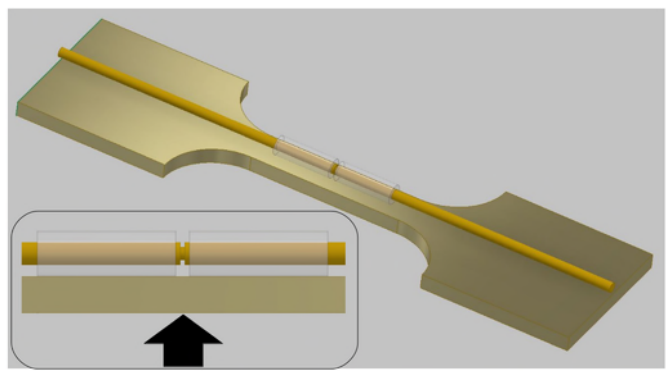

***Figure S2*** *: Schematic of the sample used for our MCBJ. The samples are made from Au wire glued into two quartz capillaries which are then glued onto a phosphorous bronze bending beam. As seen in the inset, the wire is notched in the empty space between the capillaries to initiate the breaking of the junction. See text for more details.*

Figure S2 shows a scheme of the samples used for our MCBJ measurements. They are made from Au wire (250 μm diameter, 99.99%,

Goodfellow) glued into two quartz capillaries (fused silica, 1mm inner diameter, Vitrocom) which are then glued onto a phosphorous bronze bending beam. An optical glue (NOA61, Nordland) is used for both bondings. The wire is notched in the empty space between the capillaries to initiate the breaking of the junction. The free-standing part of the Au wire is typically 200 μm for our samples, and the unfilled parts of the capillaries act as a reservoir for the organic solutions. Before use, samples are cleaned in plasma cleaner (ATTO, Diener Electronic) operated with an air pressure of 0.4 mbar at a power of 30W for two minutes.

The separation of the electrodes is controlled by a micrometer step motor (Z-825, Thorlabs) stacked up with a piezoelectric actuator (sensitivity : 216 nm.V$^{-1}$). Motors and piezo are driven through an input/output board by a dedicated computer interface, which is also used for acquiring data and feed backing. Because the wire is fixed, the actuator motion is demagnified, allowing accurate control of wire stretching. Taking into account a typical push:stretch ratio of 20:1 and the resolution of our 16-bit DAC, one digit corresponds to less than 3 pm, which is ample for this work. As the stability of the MCBJ is a critical parameter for this study, it is operated in the basement of the laboratory on an optical table to ensure optimal isolation from mechanical vibrations in a temperature-controlled environment (temperature variations below 1°C on a 24-hour scale). At low bias ($35 \leq V_{bias} \leq 140$ mV for this study) and at room temperature, the drift of the electrodes is below 5 pm.s$^{-1}$ after one or two hours of operation.

The molecular junction conductance is derived from the current intensity that flows through the junction, measured using a home-made current/voltage converter with a logarithmic trans-conductance gain following the design proposed by U. Dürig et al.[4] This converter allows measurements from the 100pA range (noise level of 10pA on a 10 kHz bandwidth) up to the mA range, and it is operated at a constant temperature of 20°C. Prior to measurements, the converter is carefully calibrated with a series of precision resistors (1,10,100 M$\Omega$ and 1 G$\Omega$ 0.1% resistors). These calibration data are then used to calculate the current or conductance values corresponding to the output voltage. The junction, in series with a 1 k$\Omega$ ballast resistor to avoid saturation of the trans-impedance amplifier, is biased using a 6V lead battery. The voltage output of the trans-impedance amplifier is recorded using a 16-bit ADC, operated at a sampling frequency of 4096 Hz. Fresh millimolar solutions of molecules in DMSO are prepared and immediately used for measurements.

***Measurement and analysis protocol.***

To ensure cleanliness of the contacts, samples are broken in pure solvent (here DMSO). They are operated for one or two hours at a typical current set point of 200 pA before measurements begin, to allow mechanical relaxation of the bending beam, and thus stabilization of the electrode distance. We then feed the sample with 10 μL of solution and set a current set point in the sub nA range. This current imposes the distance between the gold electrodes, and after stabilization, the feedback loop controlling the electrode separation is disabled. We then observe the temporal evolution of the current flowing through the contact. A connection or disconnection of a molecule between the 2 electrodes is expected to induce a sharp change (stepwise) in the current. If we do not observe such events, the feedback is enabled, and the set point is increased (reducing electrode separation). These operations are repeated until we observe abrupt current jumps. Taking advantage of thermal diffusion and electromigration, at room temperature, the nano-contact self-organizes at atomic level and naturally explores the most stable configurations around the average chosen conductance value. During these periods we record the temporal evolution of the current flowing through the contact. We operate the piezo actuator only if the current falls below 100 pA or rise above 100nA. The recording of the current is interrupted when the piezo actuator is active. Figure S3 illustrates a 8s recording of the current flowing through a junction filled with a millimolar solution of molecule **B** in DMSO, operated at a bias voltage of 35 mV. This recording exhibits a random telegraphic signal associated to numerous molecule connection/disconnection events in the contact, as demonstrated in a previous publication.[5] Current histogram clearly illustrates the discrete nature of current jumps and the most likely values of the current. Such histograms presented below are all constructed using the rule suggested by D.P. Doane for data binning,[6] in order to be able to compare histograms constructed from datasets of different sizes.

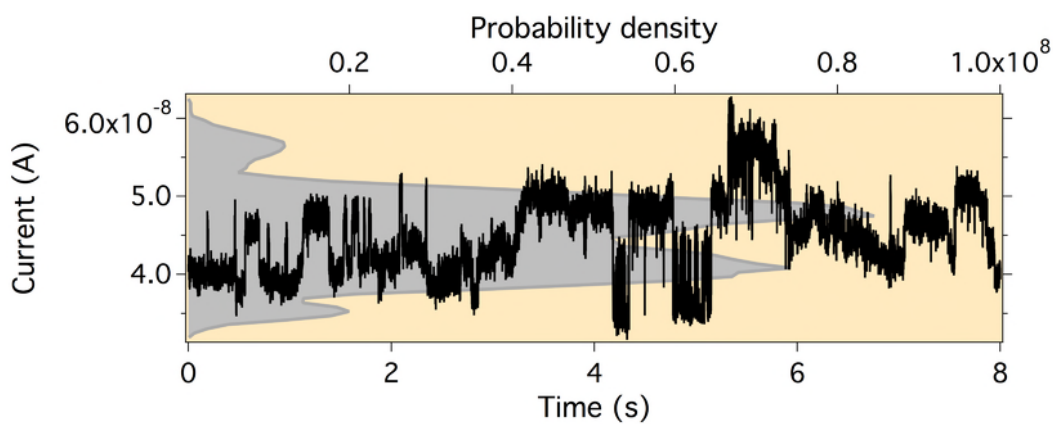

*Figure S3* : *Spontaneous evolution of the current of a molecular contact formed from a millimolar solution of molecule **B** in DMSO. A 35 mV bias is applied to the junction. The electrode separation is fixed with a set-point current of 1 nA before the regulation feedback is disabled (see text for more details). Current (black line) is recorded. A histogram of the current (gray) clearly indicates the most probable current values.*

## III. XPS measurements.

SAM molecule **A**

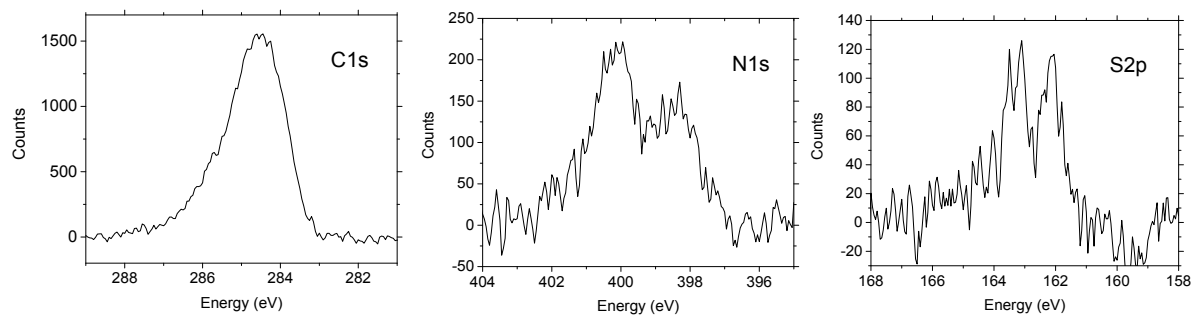

SAM molecule **B**

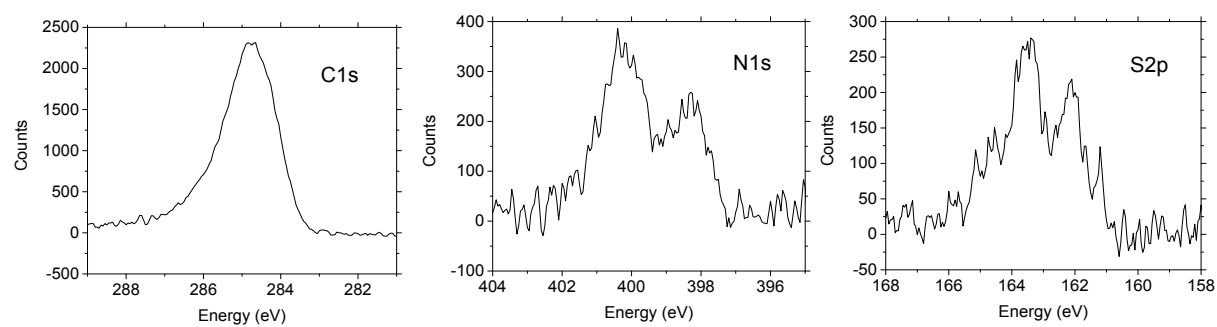

***Figure S4***. *C₁ₛ, S₂ₚ₃/₂ and N₁ₛ spectra of SAMs **A** and **B**.*

|  | *C1s* | *S2p₃/₂ (\*)* | *N1s* | *S/C* | *N/C* |
|---|---|---|---|---|---|
| *SAM A* | *284.4* | *162.1, 163.2* | *400.2, 398.4* | *0.05 (0.06)* | *0.11 (0.18)* |
| *SAM B* | *284.8* | *162.1, 163.7* | *400.4, 398.3* | *0.09 (0.1)* | *0.15 (0.2)* |

***Table S1***. *Peak position (eV) of C₁ₛ, S₂ₚ₃/₂ and N₁ₛ and atomic ratios of elements (expected values from the molecule structures in brackets).*

As expected, the S2p spectra of SAMs **A** and **B** (Fig. S5) were composed by 2 contributions of equal intensities (each of them constituted by the 2 components S2p₃/₂ and S2p₁/₂, Δ = 1.16eV, intensity ratio = 0.51). The first one located at low binding energy (162.1eV) was associated to S-Au bond, the second one at higher binding energy corresponding to the free thiol end-group.

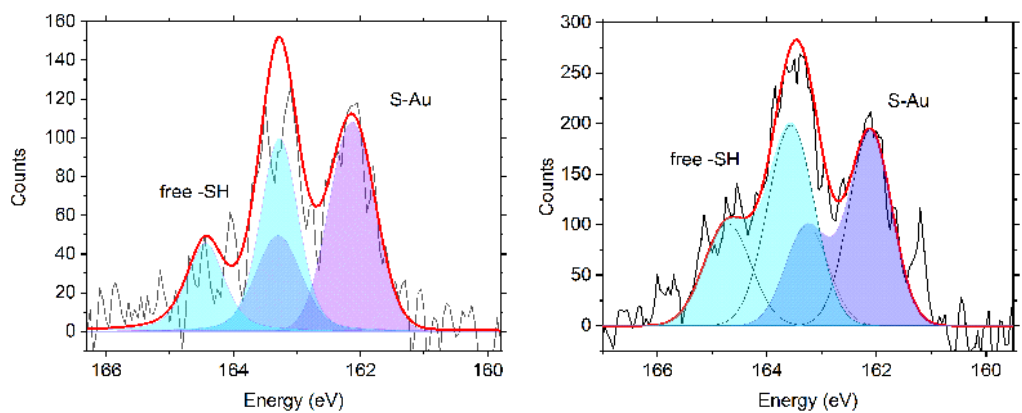

***Figure S5**. Deconvolutions of S2p spectra for SAMs **A** (left) and **B** (right).*

The interpretation of N1s spectra (Fig. S6) for SAMs **A** and **B** is more delicate. Usually 2 signals in a 1:1 intensity ratio are expected for each imidazole rings attributed to pyridinic (=N–) and pyrrolic (–NH–) nitrogens.[7-8] In the case of anilines, it is not rare to observe the presence of hydrogen-bonds and/or the protonated form Ar-$NH_2$/$NH_3^+$ in the 401-402eV region.[9-10] Here, for the SAM **A**, 3 bands are observed at 398.4eV, 400.2eV, and 401.7 eV corresponding to imine (–N=C), tertiary amine (–N<) and protonated/H-bonded nitrogens (–$NH_2$/$NH_3^+$), respectively. A ratio of 4 (–N< + $NH_2$/$NH_3^+$) for 2.2 (–N=C) is measured, which is closed to the expected value (i.e. 4 amine-type N for 2 imine-type N), knowing that photodamaging by X-rays analysis of amine terminal group into unsaturated imine adducts is also possible.[11]

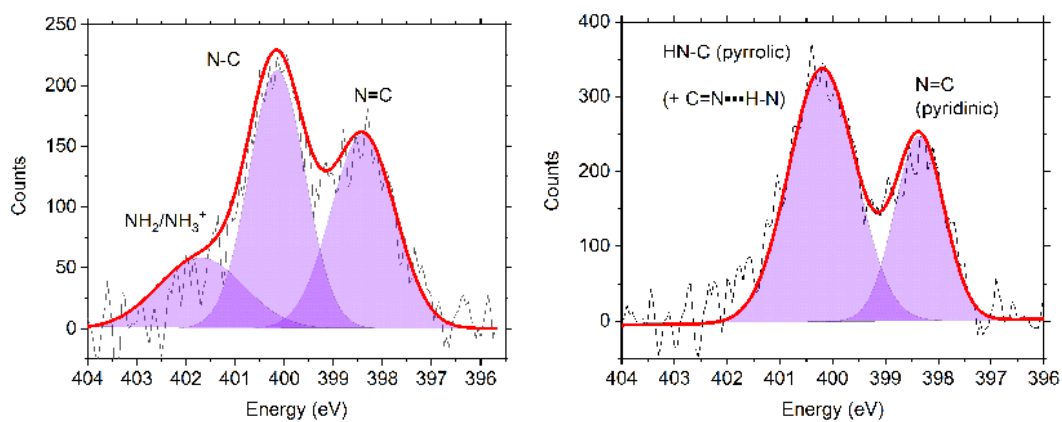

***Figure S6**. Deconvolutions of N1s spectra for SAMs **A** (left) and **B** (right).*

In SAM **B**, the N1s signal has 2 unequal components (1:0.54 intensity ratio instead of the expected 1:1 ratio) observed at 398.3eV (pyridinic N) and 400.4eV (pyrrolic N). In the literature, N1s signals with unequal areas were also observed in polybenzimidazole polymers[12-13] or in graphene functionalized by benzimidazole.[14] An explanation could be a difference in X-ray sensitivity of pyridinic nitrogens compared to pyrrolic nitrogens. Another hypothesis to explain the preponderance of the pyrrolic form would be the presence of hydrogen bonding between the pyridinic and pyrrolic nitrogens (C=N•••H–N) or due to H-bonds with residual water molecules included in the monolayer (C=N•••H–O).[15] Due to this complexity of N1s spectra for SAM **A** and **B**, it was not possible to clearly identify the signals observed in the presence of acid and base.

## IV. Measurements on <sup>TS</sup>Au/SAM/C-AFM tip junctions.

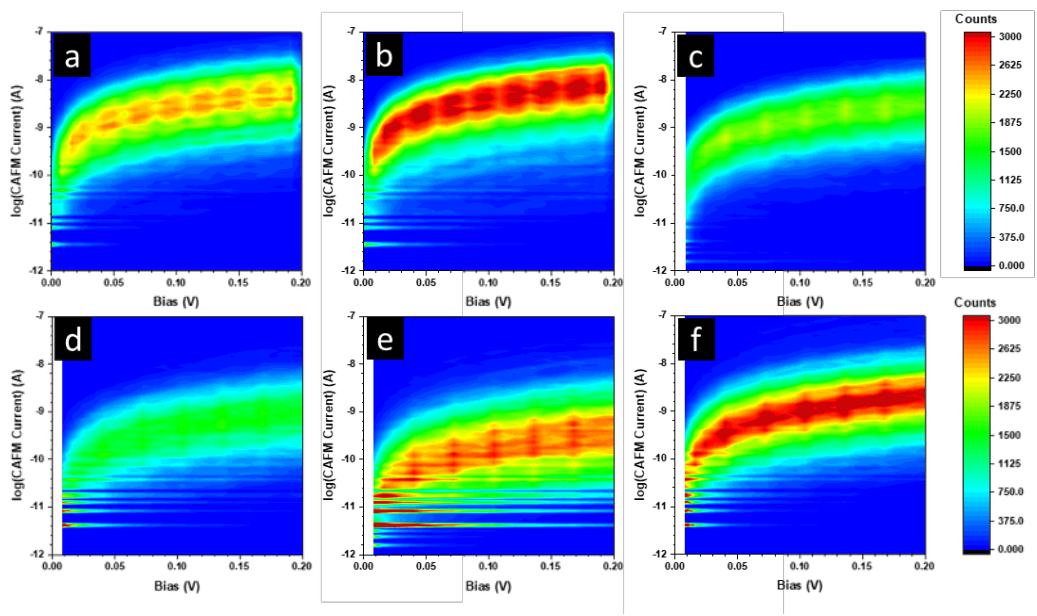

***Figure S7***. *2D histograms of the current-voltage curves (I-V) for <sup>TS</sup>Au/SAM/C-AFM tip for **(a)** pristine SAM **A** (4515 traces) **(b)** neutral SAM **A** (5403 traces); **(c)** protanated SAM **A** (3802 traces); **(d)** pristine SAM **B** (3492 traces) **(e)** neutral SAM **B** (7676 traces) and **(f)** protanated SAM **B** (6752 traces).*

| | NMJs | | $^{TS}$Au/SAMs | |
|---|---|---|---|---|
| | log μ (log A) | log σ | log μ (log A) | log σ |
| **mol. A - pristine** | n.m. | n.m. | -8.32 ± 0.02 (8.51x10$^{-9}$ A) | 0.57 ± 0.05 |
| **- protonated** | -8.07 ± 0.02 (8.51x10$^{-9}$ A) | 0.33 ± 0.02 | -8.81 ± 0.01 (1.55x10$^{-9}$) | 0.36 ± 0.01 |
| **- neutral** | -7.54 ± 0.01 (2.88x10$^{-8}$ A) | 0.05 ±0.01 | -7.86 ± 0.01 (1.38x10$^{-8}$) | 0.69 ± 0.02 |
| **mol. B - pristine** | n.m. | n.m. | -9.08 ± 0.01 (8.32x10$^{-10}$ A) | 0.59 ± 0.02 |
| **- protonated** | -7.38 ± 0.01 (4.17x10$^{-8}$ A) | 0.29 ± 0.01 | -8.69 ± 0.01 (2.04x10$^{-9}$ A) | 0.50 ± 0.01 |
| **- neutral** | -7.78 ± 0.02 (1.66x10$^{-8}$ A) | 0.51 ± 0.03 | -9.51 ± 0.02 (3.09x10$^{-10}$ A) | 0.69 ± 0.02 |

***Table S2***. *Fitted parameters (mean values and standard deviations) of the log normal distributions for NMJs and $^{TS}$Au/SAMs junctions. In brakets: the mean current in A. (n.m. =not measured).*

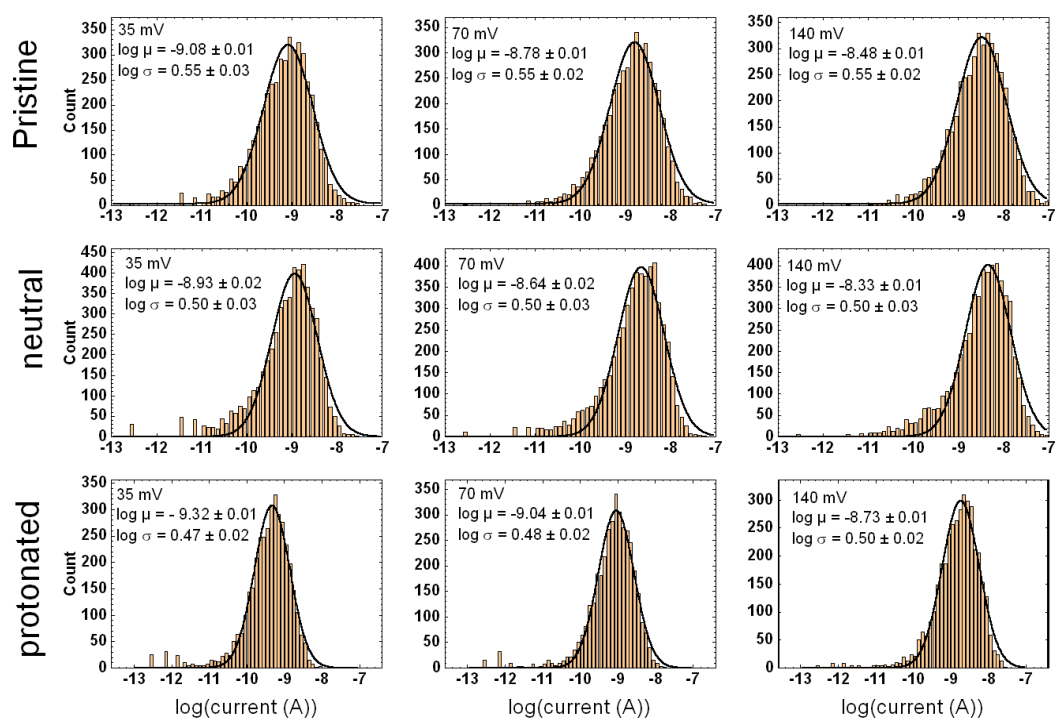

**Figure S8.** *Current histograms of the SAM of molecule **A** at 35, 70 and 140 mV (from data in Fig. S7a-c) for the pristine, protonated and after converted back to the neutral state.*

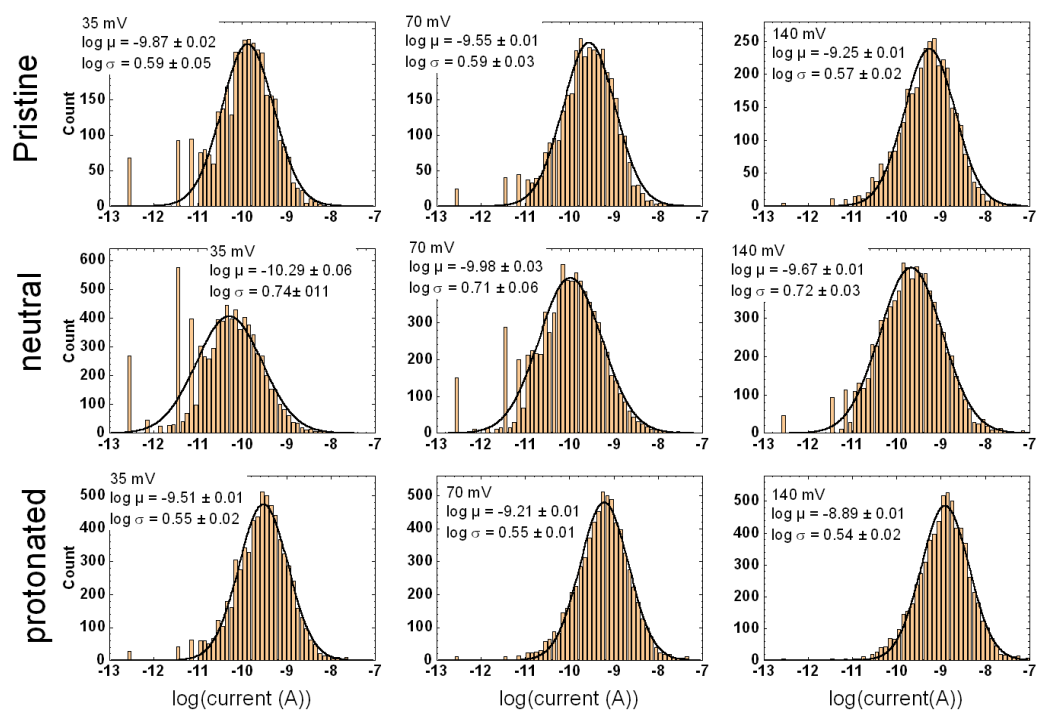

**Figure S9.** *Current histograms of the SAM of molecule **B** at 35, 70 and 140 mV (from data in Fig. S7a-c) for the pristine, protonated and after converted back to the neutral state.*

| Bias (mV) | A pristine | A protonated | A neutral | B pristine | B protonated | B neutral |
|-----------|------------|--------------|-----------|------------|--------------|-----------|
| **35** | $3.1 \times 10^{-4}$ | $1.8 \times 10^{-4}$ | $4.2 \times 10^{-4}$ | $5.0 \times 10^{-5}$ | $2.1 \times 10^{-4}$ | $1.9 \times 10^{-5}$ |
| **70** | $3.0 \times 10^{-4}$ | $1.7 \times 10^{-4}$ | $4.2 \times 10^{-4}$ | $5.1 \times 10^{-5}$ | $1.1 \times 10^{-4}$ | $2.0 \times 10^{-5}$ |
| **140** | $3.1 \times 10^{-4}$ | $1.7 \times 10^{-4}$ | $4.2 \times 10^{-4}$ | $5.2 \times 10^{-5}$ | $1.1 \times 10^{-4}$ | $5.7 \times 10^{-5}$ |
| **200** | $3.1 \times 10^{-4}$ | $9.7 \times 10^{-5}$ | $9.0 \times 10^{-4}$ | $5.4 \times 10^{-5}$ | $1.3 \times 10^{-4}$ | $2.1 \times 10^{-5}$ |

**Table S3.** *Conductances fo the SAMs (in unit of $G_0$) at 35, 70, 140 and 200 mV for molecules **A** and **B** for the pristine, protonated and after converted back to the neutral state (conductances are the mean currents from the peaks of the current histograms - Figs 1, S8 and S9, divided by the applied voltage).*

## V. C-AFM on NMJs.

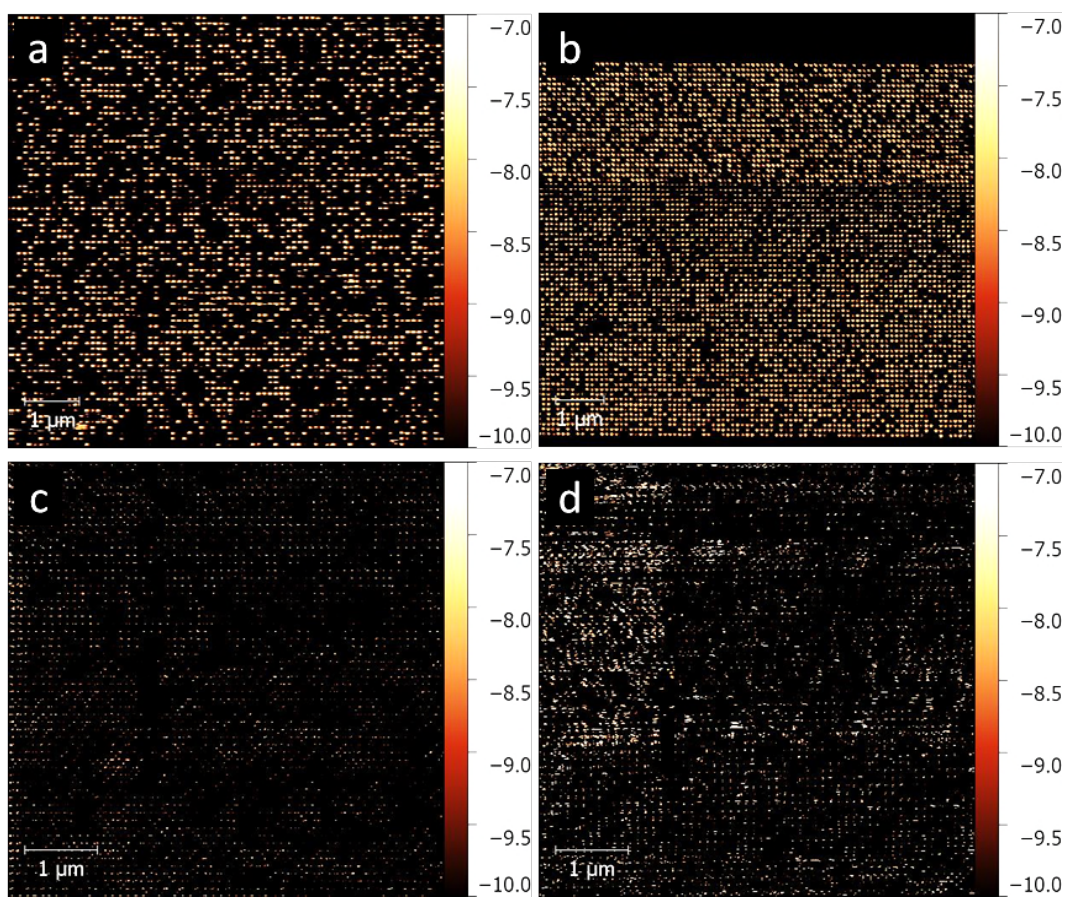

***Figure S10***. *Current image (bright spots are the current value of each NMJ - the scale is in log of the current measured in ampere) obtained by C-AFM (loading force 10 nN and $V_{bias} = 200$ mV) on NMJs with **(a)** neutral molecules **A** (2456 nanodots); **(b)** protanated molecules **A** (2445 nanodots); **(c)** neutral molecules **B** (1154 nanodots) and **(d)** protanated molecules **B** (1634 nanodots).*

## VI. MCBJ measurements.

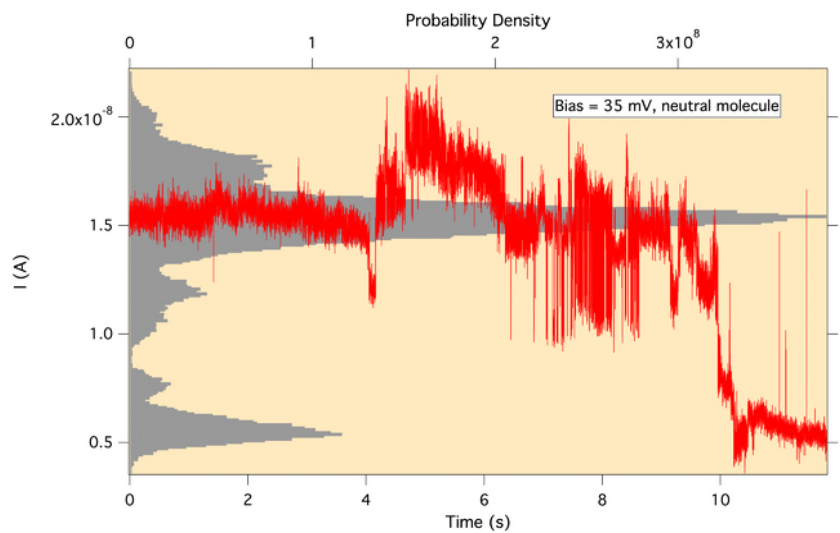

***Figure S11***. *Current vs. time evolution (red) and corresponding current histogram (grey) of molecular junction formed from a millimolar solution of molecule **B** (neutral) in DMSO. A 35 mV bias is applied to the junction. The electrode separation is fixed with a set-point current of 1 nA before the regulation feedback is disabled (see text for more details). Current (black line) is recorded. A histogram of the current (gray) clearly indicates the most probable current values.*

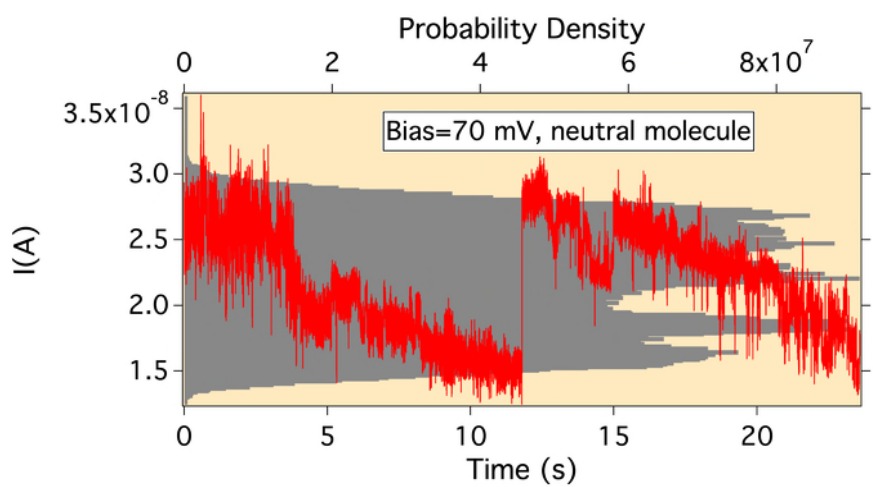

**Figure S12**. *Current vs. time evolution (red) and corresponding current histogram (grey) of molecular junction formed from a millimolar solution of molecule **B** (neutral) in DMSO at $V_{bias}$ = 70 mV.*

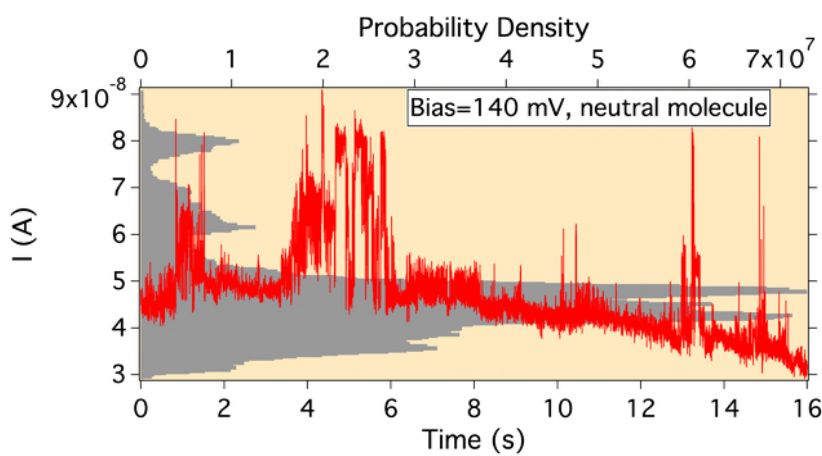

**Figure S13**. *Current vs. time evolution (red) and corresponding current histogram (grey) of molecular junction formed from a millimolar solution of molecule **B** (neutral) in DMSO at $V_{bias}$ = 140 mV.*

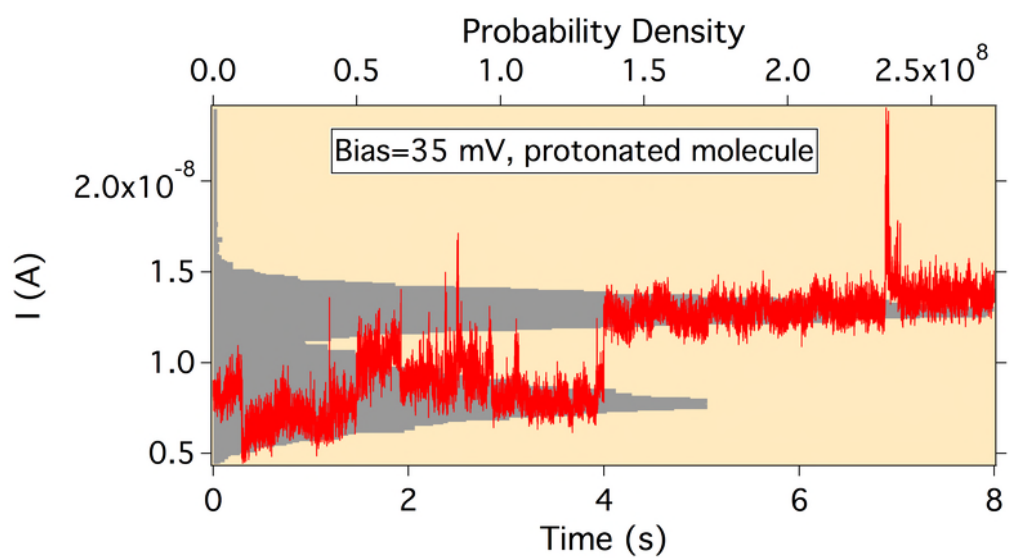

**Figure S14**. *Current vs. time evolution (red) and corresponding current histogram (grey) of molecular junction formed from a millimolar solution of molecule B (protonated) in DMSO at $V_{bias}$ = 35 mV.*

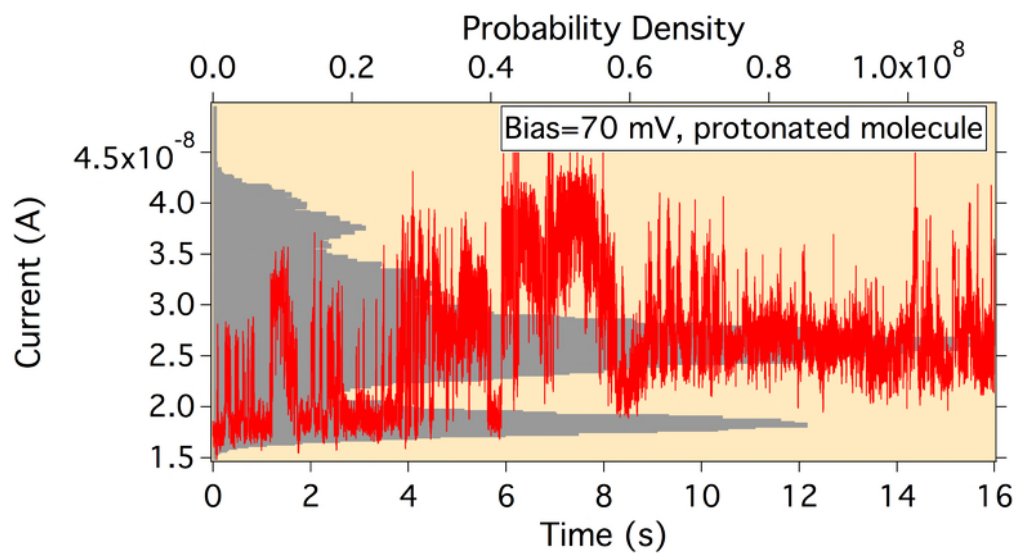

**Figure S15**. *Current vs. time evolution (red) and corresponding current histogram (grey) of molecular junction formed from a millimolar solution of molecule B (protonated) in DMSO at $V_{bias}$ = 70 mV.*

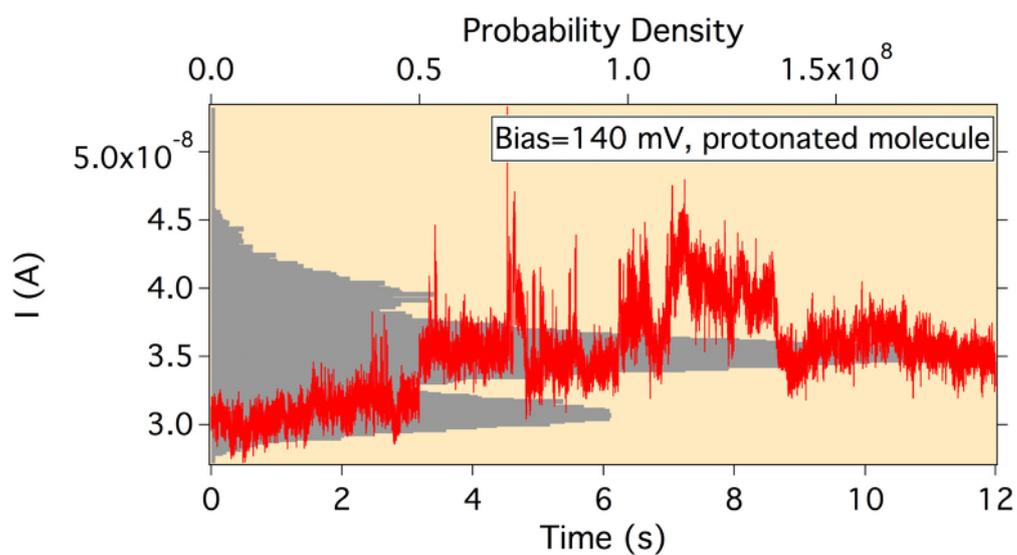

***Figure S16***. *Current vs. time evolution (red) and corresponding current histogram*
*(grey) of molecular junction formed from a millimolar solution of molecule **B***
*(protonated) in DMSO at $V_{bias}$ = 140 mV.*

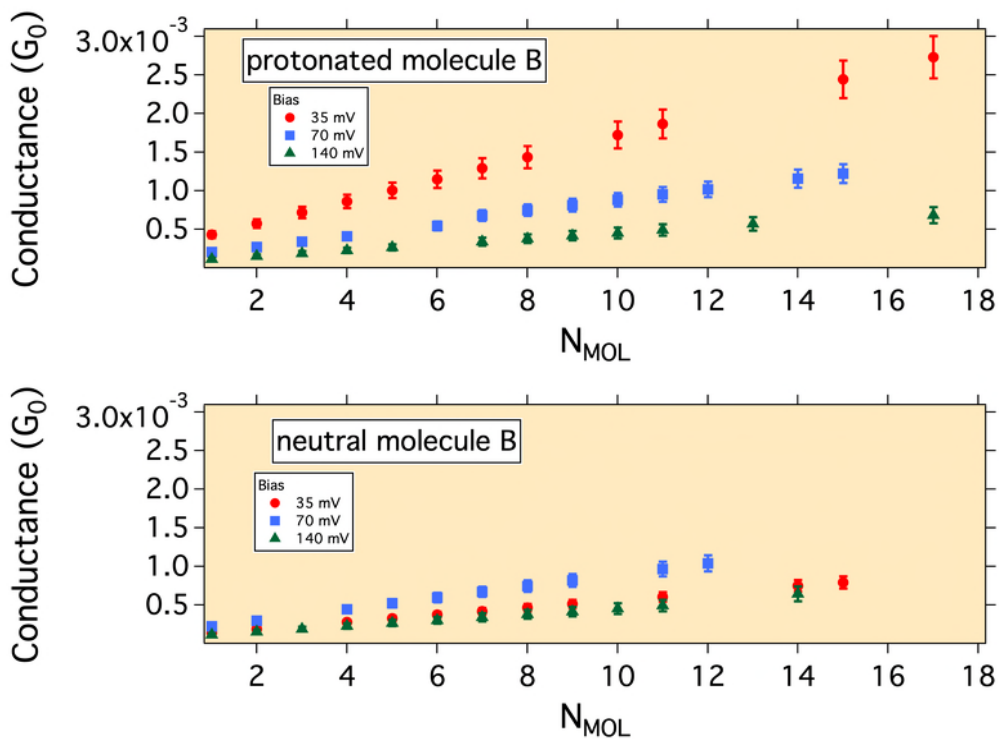

**Figure S17**. *Conductance values of the peaks observed in the histograms of currents versus a number of molecules (up panel: protonated molecule **B** from Figs. S14-S16, down panel: neutral molecule **B**, from Figs. S11-S13). A linear fit gives the $G_{MOL}$ values reported en table 2 (main text) for the 3 applied voltages. Conductance values are given in conductance quantum units, $G_0$.*

## VII. Modeling, molecule in gas phase

For the modeling of the protonation, we used a cluster model where the protonation reaction for the molecule **A** or **B** is empirically defined by:[16]

A + nHCl → A●nH⁺ + nCl⁻ 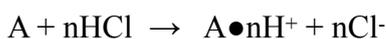

In this cluster approach, the charge transfer is partially implemented (the Mülliken charge estimated by ab-initio calculation of the chlorine atom is -0793e and 0.330e for the hydrogen) and this is clearly a   physical limit compared to the simple chemical picture sketched by the previous equation. To model the protonated molecules, an HCl molecule has been introduced close to the nitrogen atoms corresponding to the most plausible protonation sites, and the protonated states were minimized in terms of total energy.

*- Molecule geometry optimizations*

Figure S14 shows the optimized geometries for molecules **A** and **B**. The center of the molecule (imidazole) and the two phenyl groups remain planar. The length of both molecules **A** and **B** is 18.9Å. The major difference is observed for the lateral size of the two molecules : 5.1Å for **A** and 16.3Å for **B**. For the unprotonated state of molecule **A** the dihedral angle between the alanine rings and imidazole is near 90°. For both molecules, the imidazole ring presents the most accessible protonation sites with the presence of two non-hydrogenated nitrogen atoms. The introduction of a HCl molecule leads to the formation of N-H bond characterized by a bond length of 1.0764 Å. The N-H bond length is characteristic of a covalent bond. The protonation leads to a H-Cl distance of 1.891 Å larger than the HCl bond of 1.318Å. For the molecule **A**, up to four protonation degrees can be foreseen with the protonation of the nitrogen atoms belonging to the alanine ring as shown in Figure S14 without disturbing the global geometry of the molecule.

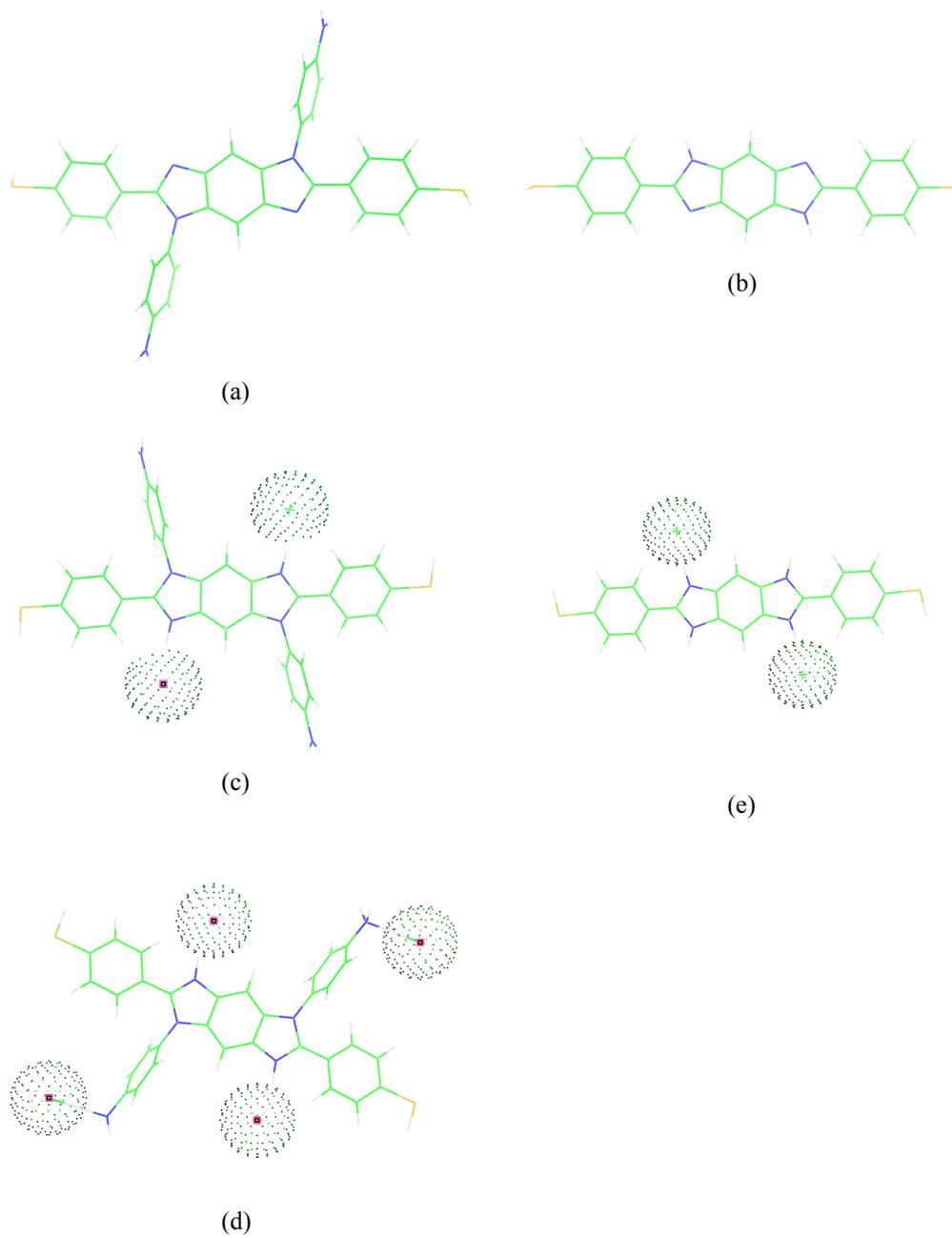

(a)

(b)

(c)

(e)

(d)

***Figure S18.*** *Optimized geometries of **(a)** unprotonated molecule **A**, **(b)** unprotonated molecule **B**, **(c)** protonated molecule **A** with two protons and **(d)** 4 protons, **(e)** protonated molecule **B** with two protons.*

*- Energy levels and molecular orbitals*

Figures S15 show the evolution of the HOMO-LUMO gap, the EA and IP of the molecules **A** and **B** as function of the protonation states. For the molecule **A**, most of the HOMO-LUMO band gap variation is obtained when 2 protons are on the imidazole. Protonation of the amine groups have no great effect (see also the molecular orbitals, Figure S16). Basically, the protonation tends to reduce the HOMO-LUMO gap of the molecules and increases the IP and EA. Both the HOMO and the LUMO of molecules **A** and **B** are characterized by highly delocalized wave functions (Figure S16). The protonation of the nitrogen atoms strongly modify the nature of the HOMO as it becomes strongly localized on the protonation sites i.e. on the chlorine counter-ion (Figure S16). On the contrary, the LUMO wave functions remain strongly delocalized on the aromatic cycles.

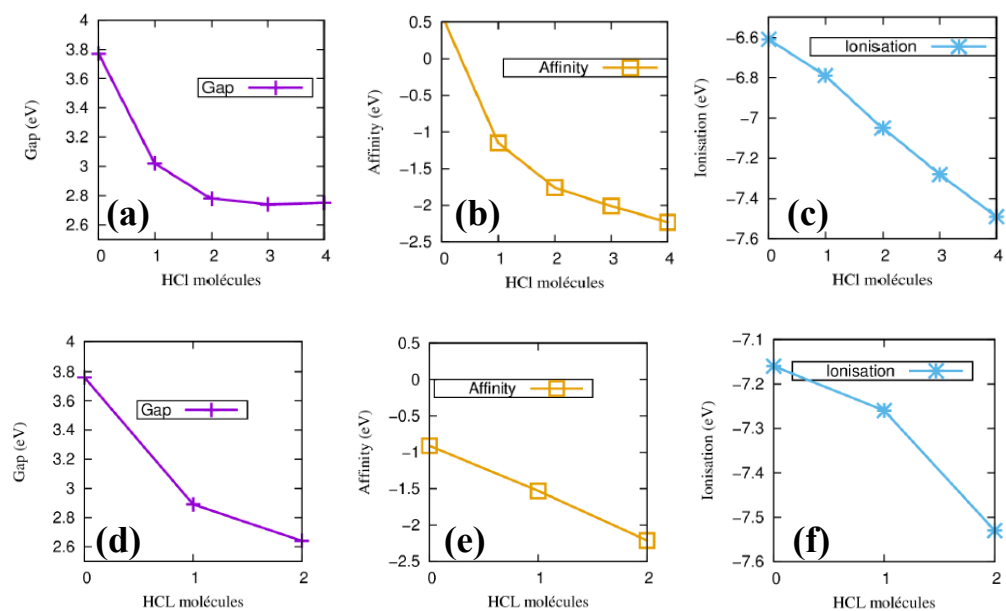

***Figure S19***. *Evolution of **(a)** the HOMO-LUMO gap, **(b)** electron affinity and **(c)** ionization potential of molecule **A** versus the protonation state. Evolution of **(d)** the HOMO-LUMO gap, **(e)** electron affinity and **(f)** ionization potential of molecule **B** versus the protonation state.*

| | HOMO | LUMO |
|---|---|---|
| **molecule A** | 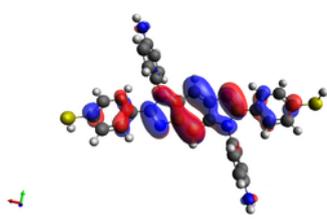 | 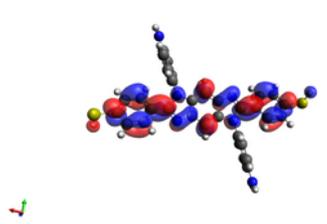 |
| **molecule A + 2H⁺** | 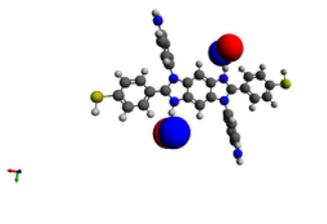 | 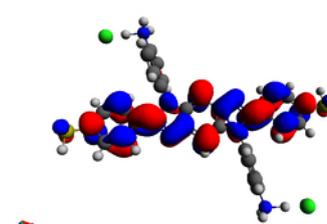 |
| **molecule A + 4H⁺** | 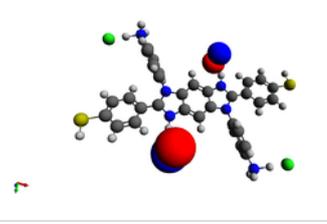 | 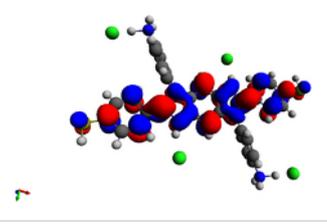 |
| **molecule B** | 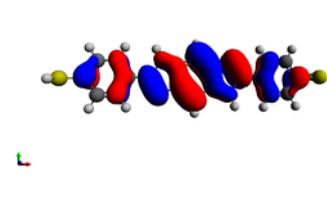 | 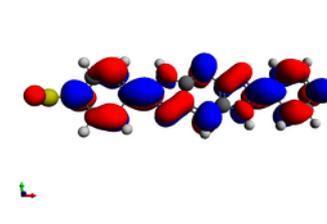 |
| **molecule B + 2H⁺** | 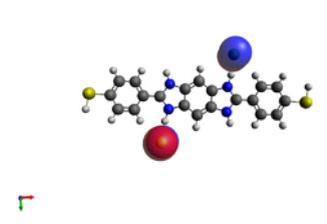 | 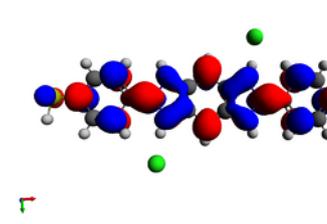 |

***Figure S20**. HOMO and LUMO wave functions.*

*- Optical properties*

The optical properties of the two molecules were calculated by Time Dependent Density Functional Theory (TDDFT) calculations using the B3LYP functional. Only vertical transitions are considered and the activate space considered correlates 12 electrons with 12 orbitals. For the molecule **A** (respectively **B**), the major optical transition takes place at 346 nm (357 nm, respectively) in good agreement with the measurements of the precursor (Figure S1) with an oscillator strength of 0.9 (respectively 1.28). As shown in Figure S17, upon protonation, a bathochromic shift to 387 nm (385 nm, respectively), i.e. an increase of 41 nm (respectively 28 nm) is observed in qualitative agreement with the experimental observation (Figure S1). This corresponds to a relatively small reduction of the optical gap for molecule **A** (respectively **B**) which is reduced from 3.58 eV to 3.2 eV (respectively from 3.47 to 3.2 eV).

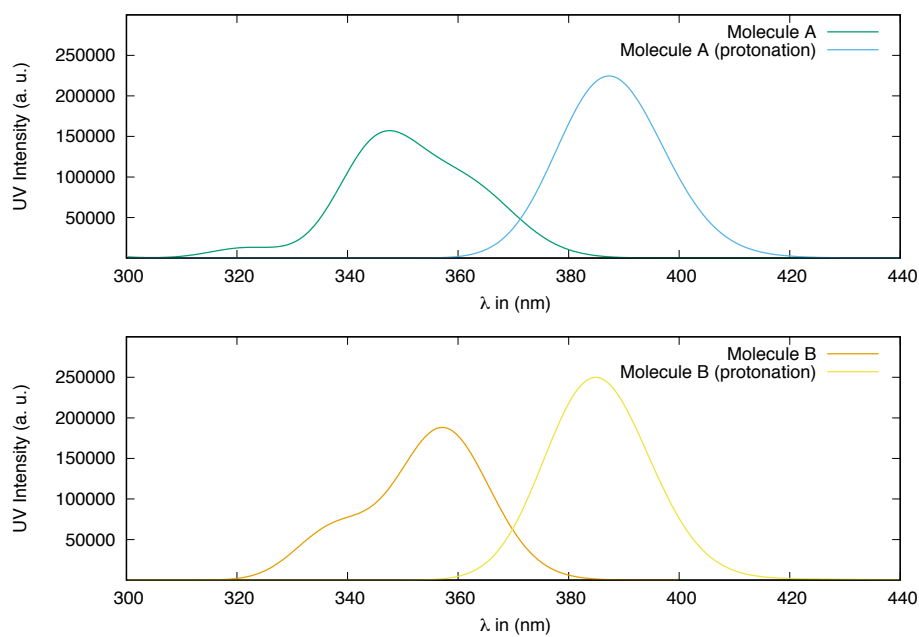

**Figure S21**. *TDDFT simulation of the optical spectrum for molecules **A** and **B** with and without protonation. The calculations indicate that the introduction of localized states near the HOMO degrades the intensity of the optical oscillator*

*and the activated occupied states. The active optical transition would be HOMO-6→ LUMO. Protonation states considered are the highest possible for the two molecules (i.e. 4 for **A** and 2 for **B**).*

## VIII. Modeling electron transport in metal/molecule/metal junction.

To model the behavior of a metal/molecule/metal junction, we attach the geometry-optimized molecule to gold via the thiol anchor groups. In this case we bind the terminal sulfur atom to a hollow site on a (111) gold surface and the optimum Au-S distance is calculated to be 2.3Å. The hydrogen atom contacted to the sulfur is removed. An extended molecule is then constructed to consist of 6 layers of (111) gold each containing 25 atoms. Using the DFT code SIESTA,[17] a Hamiltonian describing the extended molecule is generated using the following parameterization. A double-ζ plus polarization basis set, energy cut off=150 Rydbergs, norm conserving pseudopotentials, and the GGA method[18] to describe the exchange correlation. The zero bias transmission coefficient T(E) is then calculated using this Hamiltonian, via the GOLLUM code.[19] The room temperature conductance can then be evaluated from the Landauer formula,

$$G = \frac{2e^2}{h} \int_{-\infty}^{\infty} dE \, T(E) \left( \frac{-df(E)}{dE} \right)$$

Where, *f(E)* is the Fermi-Dirac distribution, *e* is the electronic charge and *h* is Planck's constant. The electrical current is evaluated using the equation,

$$I = \frac{2e}{h} \int_{-\infty}^{\infty} dE \, T(E) \left[ f_L(E) - f_R(E) \right]$$

Where $f_L$ and $f_R$ are the Fermi-Dirac distributions in the left and rights lead respectively.

### -Conductance of neutral molecules A and B

Figure S18 shows the calculated conductance through molecules **A** and **B** for the geometry given in Figure 6. Here the HOMO resonances of the two molecules lie at similar positions relative to the SIESTA predicted Fermi Energy $E_F^0$. This is in disagreement with the calculated behavior in the gas phase (Fig. S15), where the HOMO level of **B** lies at a lower energy. This can be attributed to the thiol anchor groups (after losing a hydrogen) pinning the HOMO close to the Fermi energy in this calculation.

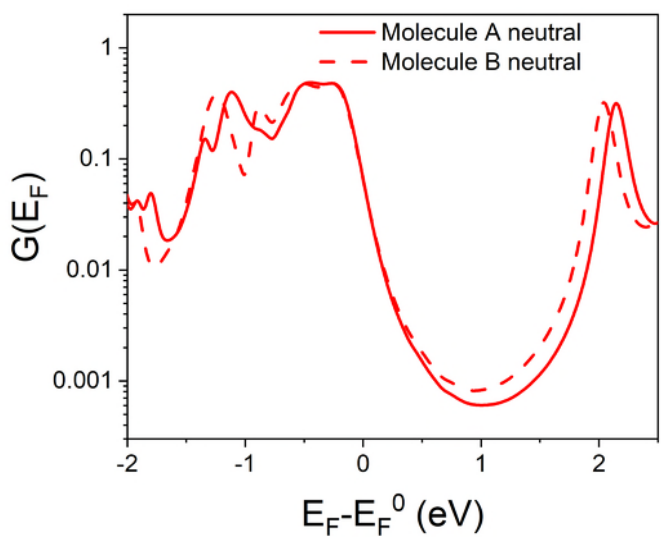

***Figure S22***. *Electrical conductance of the deprotonated molecules **A** and **B***.

### -Electron transport through a protonated molecule

Protonation of the molecule, by adding hydrogen atoms at the nitrogen sites, causes the molecule to become charged. Therefore, counter ions, which here are chlorine are needed to balance the charge i.e. making the added hydrogen atoms

H+ and the chlorine atoms Cl-. To achieve this within SIESTA requires analysis of the electron distribution using a Mulliken population analysis. To form a Cl- ion requires the number electrons on the chlorine atom to increase by one (here we define N as the number of electrons added to the chlorine atom). The charge on the chlorine atom can be controlled by varying the cut-off radius in the basis set. Here we set the Cl-H distance to be 3Å. Figure S19 shows the transmission for the protonated molecule B, for different values of N on the counter ions. When there are no counter ions, the LUMO resonance lies directly at the Fermi energy as expected for a charged system (cyan line). The addition of the counter ion shifts the position of the Fermi energy towards the middle of the HOMO-LUMO gap and the extent of the shift is controlled by the value of N on the counter ions. Here N=0.7 (blue line) and N=1.0 (black line).

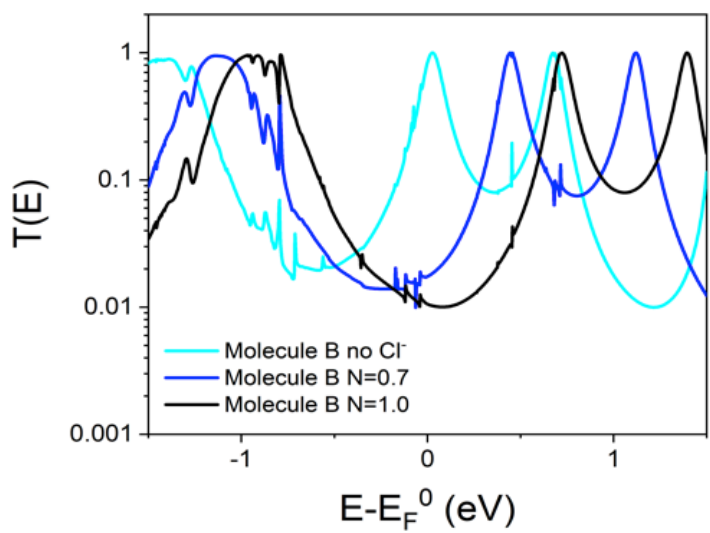

***Figure S23****. Transmission coefficient T(E) for the protonated molecule **B** when they are no Cl- counter ions (cyan), when the number of electron N on the Cl atom =0.7 (blue) and when N=1 (black).*

**-Tilt angle**

The SAM of molecule **A** has a smaller thickness than **B**, which suggests the molecule is tilted. Figure S20 shows the calculation for a tilt angle of 60° relative to the surface normal (the gold-gold separation is now 1.4nm). The resulting behavior is similar to that for the linear geometry in figure 6, with the transmission decreasing when the molecule is protonated.

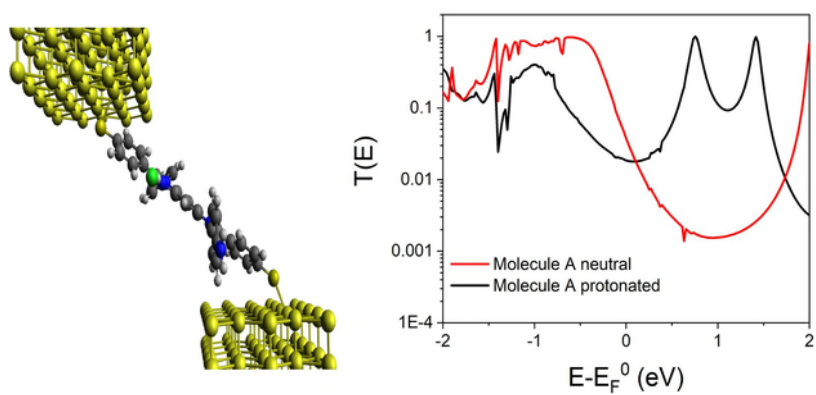

***Figure S24***. *Zero bias transmission coefficient T(E) for molecule **A** tilted at an angle of 60° away from the normal.*

**-*Transport through NH₂ anchor groups in Molecule A***

Molecule **A** contains $NH_2$ groups, which may couple to the gold electrodes. We calculate for this geometry, where the N-Au distance is 2.4 (figure S21). The resulting transmission shows that the off-resonance values are much lower than the equivalent molecule binding through the thiols. At the Fermi energy the value of is approximately 3 orders of magnitude lower. This is due to the ring rotation, which is approximately 90° to the central core, and suggest that these groups play a minimal role in the electron transport.

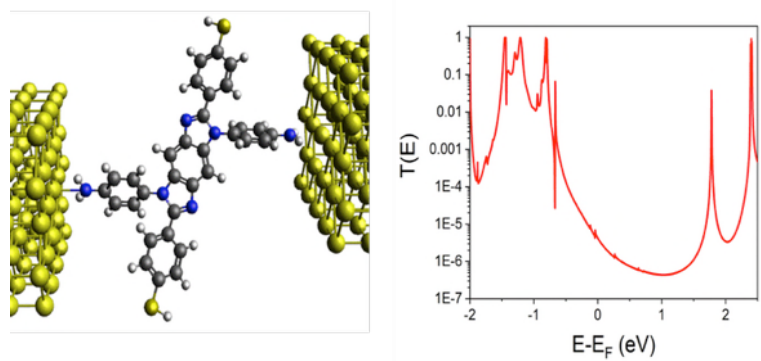

**Figure S25.** *(left) Geometry of molecule **A** contacted to gold electrodes through the NH₂ groups and (right) the corresponding zero bias transmission coefficient T(E).*